\documentclass[11pt]{article}

\usepackage{graphics} % for pdf, bitmapped graphics files
\usepackage{psfig} % for postscript graphics files
\usepackage{times} % assumes new font selection scheme installed
\usepackage{amsmath} % assumes amsmath package installed
\usepackage{amssymb}  % assumes amsmath package installed

\newtheorem{theorem}{Theorem}
\newtheorem{lemma}{Lemma}[section]
\newtheorem{proposition}[lemma]{Proposition}
\newtheorem{corollary}[lemma]{Corollary}
\newtheorem{fact_}{Fact}

\newenvironment{theo}[1]{\begin{theorem}\label{#1}\rm}{\end{theorem}} 
\newenvironment{lem}[1]{\begin{lemma}\label{#1} \rm}{\end{lemma}} 
 
\newenvironment{cor}[1]{\begin{corollary}\label{#1}\rm}{\end{corollary}}

\newcommand{\beqn}[1]{\begin{eqnarray}\label{#1}}
\newcommand{\eeqn}{\end{eqnarray}}
\newcommand{\beq}{\begin{eqnarray*}}
\newcommand{\eeq}{\end{eqnarray*}}
\newcommand{\bit}{\begin{itemize}}
\newcommand{\eit}{\end{itemize}}
\newcommand{\rf}[1]{~(\ref{#1})}
\newcommand{\comment}[1]{}
\newcommand{\bprf}{{\it Proof: }}
\newcommand{\bbox}{\hfill\rule{1ex}{1.4ex}}

\newcommand{\eps}{\varepsilon}
\newcommand{\N}{{\mathbb N}}

\newcommand{\tProt}{\mbox{{\tiny\it Prot}}}
\newcommand{\tmRNA}{\mbox{{\tiny\it mRNA}}}
\newcommand{\cX}{\widehat{\mbox{\it X}}}
\newcommand{\cwg}{\widehat{\mbox{\it wg}}}
\newcommand{\cWG}{\widehat{\mbox{\it WG}}}
\newcommand{\cen}{\widehat{\mbox{\it en}}}
\newcommand{\cEN}{\widehat{\mbox{\it EN}}}
\newcommand{\chh}{\widehat{\mbox{\it hh}}}
\newcommand{\cHH}{\widehat{\mbox{\it HH}}}
\newcommand{\cci}{\widehat{\mbox{\it ci}}}
\newcommand{\cCI}{\widehat{\mbox{\it CI}}}
\newcommand{\cCIA}{\widehat{\mbox{\it CIA}}}

\newcommand{\cPTC}{\widehat{\mbox{\it PTC}}}
\newcommand{\RX}{\mbox{\it RX}}
\newcommand{\slp}{\mbox{\it slp}}
\newcommand{\SLP}{\mbox{\it SLP}}
\newcommand{\wg}{\mbox{\it wg}}
\newcommand{\WG}{\mbox{\it WG}}
\newcommand{\en}{\mbox{\it en}}
\newcommand{\EN}{\mbox{\it EN}}
\newcommand{\hh}{\mbox{\it hh}}
\newcommand{\HH}{\mbox{\it HH}}
\newcommand{\ci}{\mbox{\it ci}}
\newcommand{\CI}{\mbox{\it CI}}
\newcommand{\CIA}{\mbox{\it CIA}}
\newcommand{\CIR}{\mbox{\it CIR}}
\newcommand{\PTC}{\mbox{\it PTC}}
\newcommand{\ptc}{\mbox{\it ptc}}
\newcommand{\twg}{\mbox{\tiny\it wg}}
\newcommand{\tWG}{\mbox{\tiny\it WG}}
\newcommand{\ten}{\mbox{\tiny\it en}}
\newcommand{\tEN}{\mbox{\tiny\it EN}}
\newcommand{\thh}{\mbox{\tiny\it hh}}

\newcommand{\tci}{\mbox{\tiny\it ci}}
\newcommand{\tCI}{\mbox{\tiny\it CI}}
\newcommand{\tCIA}{\mbox{\tiny\it CIA}}
\newcommand{\tCIR}{\mbox{\tiny\it CIR}}
\newcommand{\tPTC}{\mbox{\tiny\it PTC}}
\newcommand{\tptc}{\mbox{\tiny\it ptc}}

\newcommand{\lthw}[1]{\ln\, \frac{{#1}}{\theta}}   %1/\theta
\newcommand{\lthow}[1]{\ln\, \frac{{#1}}{1-\theta}}  %1/(1-\theta)
\newcommand{\lth}{\ln \frac{1}{\theta}}
\newcommand{\ltho}{\ln \frac{1}{1-\theta}}

\title{
Methods of robustness analysis for Boolean models \\of gene control networks
\footnote{M. Chaves is with the Institute for Systems Theory and Automatic Control, 
        University of Stuttgart, Pfaffenwaldring 9, 70550 Stuttgart, Germany
        {\tt\small chaves@ist.uni-stuttgart.de}.
        E.D. Sontag is with the Department of Mathematics, 
        Rutgers University, Piscataway, NJ 08854 USA
        {\tt\small sontag@math.rutgers.edu}.
        R. Albert is with the Department of Physics and Huck Institutes for the Life Sciences, 
        Pennsylvania State University, University Park, PA 16802 USA
        {\tt\small ralbert@phys.psu.edu}.
        }
}

\author{Madalena Chaves, Eduardo D. Sontag and R\'eka Albert\thanks{Corresponding author.} 
}

\date{}

\begin{document}

\maketitle

\begin{abstract}
As a discrete approach to genetic regulatory networks, Boolean models
provide an essential qualitative description of the structure of interactions
among genes and proteins. Boolean models generally assume only two possible states (expressed
or not expressed) for each gene or protein in the network as well as a
high level of synchronization among the various regulatory processes.

In this paper, we discuss and compare two possible methods of 
adapting qualitative models to incorporate
the continuous-time character of regulatory networks. 
The first method consists of introducing asynchronous updates in the Boolean 
model. In the second method, we adopt the approach introduced by L. Glass
to obtain a set of piecewise linear differential equations which continuously
describe the states of each gene or protein in the network.

We apply both methods to a particular example: a Boolean model of the 
segment polarity gene network of {\it Drosophila melanogaster}. We analyze the 
dynamics of the model, and provide a theoretical characterization of 
the model's gene pattern prediction as a function of the timescales 
of the various processes.
\end{abstract}

\section{Introduction}

Genes and gene products interact on several levels. At the genomic level, transcription 
factors can activate or inhibit the transcription of genes to give mRNAs. Since these 
transcription factors are themselves products of genes, the ultimate effect is that genes 
regulate each other's expression as part of gene regulatory networks. Similarly, proteins 
can participate in diverse post-translational interactions that lead to modified protein 
functions or to formation of protein complexes that have new roles; the totality of these 
processes is called a protein-protein interaction network.  In many cases different 
levels of interactions are integrated - for example, when the presence of an external 
signal triggers a cascade of interactions that involves biochemical 
reactions, protein-protein interactions and transcriptional regulation.

During the last decade, genomics, transcriptomics and proteomics have produced an 
incredible quantity of molecular interaction data, contributing to genome-scale maps of  
protein interaction networks~\cite{giot03,han04,li04} and transcriptional regulatory 
networks~\cite{lee02}. Network analysis of these maps revealed intriguing topological 
similarities~\cite{a05}. At the same time, it  has been increasingly realized 
that cellular interaction maps only represent a network of possibilities, and 
not all edges are present and active {\it in vivo} in a given condition or 
in a given cellular location~\cite{han04,lbyst04}.
Therefore, only an integration of (time-dependent) interaction and activity information 
will be able to give the correct dynamical picture of a cellular network.  

For many biological networks, and in particular genetic control or regulatory networks,
detailed information on the kinetic rates of protein-protein or protein-DNA 
interactions is rarely available. However, for many biological systems, 
evidence shows that regulatory relationships can be sigmoidal and be 
well approximated by step functions.
In this case, Boolean models, where every variable has only two states (ON/OFF),
and the dynamics is given by a set of logical rules, are frequently appropriate 
descriptions of the network of interactions among genes and proteins. 
Examples include models of genetic networks in the fruit fly 
{\it Drosophila melanogaster}~\cite{st01,ao03} and the flowering plant 
{\it Arabidopsis thaliana}~\cite{mta99,epa04}. 

While Boolean models introduce biologically unrealistic time constraints 
(typically, such models use synchronous updates, which inherently assume 
that the various biological processes have the same duration), 
they still provide significant qualitative information on the 
underlying structure of the network.
On the other hand, continuous models certainly have a more realistic time description of
a biological system. But, in the absence of information on the kinetic rates, continuous 
models  include many unknown parameters, and analysis of the system involves exploring
the (often large) state space of parameters. 
An important (continuous) model for {\it Drosophila melanogaster} segment polarity genes 
was first developed in~\cite{dmmo00}, where a thorough investigation of the parameter 
space showed that the system is very robust with respect to variations in the kinetic 
constants.
Both this continuous model and the discrete model~\cite{ao03} agree in their overall 
conclusions regarding the robustness of the segment polarity gene network.

In this paper, we propose two new approaches to the analysis of Boolean models,
which combine discrete logical rules and structure with more realistic 
assumptions regarding the relative timescales of the genetic processes.
The first method introduces asynchronous updates in the Boolean model; since
update times are randomly chosen, the model is now stochastic
%thereby also building an element of stochasticity into the system 
(see also~\cite{cas05}). The second method associates to the discrete variables a set of continuous variables, 
whose dynamics is given by a piecewise linear system of differential equations,
thus introducing a simple ``hybrid'' model in the manner first suggested by L. Glass and
collaborators in~\cite{g75,eg00,gk73}. 
These methods allow us to naturally probe the system with respect to perturbations in
the time dynamics to analyze its performance. 
Both methods uncover the robustness of the segment polarity gene
model~\cite{ao03}, and its ability to correctly predict the final gene 
expression pattern. 

Section~\ref{sec-drosophila} summarizes the Boolean model~\cite{ao03}, to be analyzed
with our methods. The asynchronous updates and the Glass-type methods are described 
and applied in Sections~\ref{sec-asynch} and~\ref{sec-glass}, respectively. 
In Section~\ref{sec-mutants}, we discuss the effects of perturbations in initial gene expression,
and the development of mutant phenotypes. 
Finally, in Section~\ref{sec-timesep} we show that under a timescale separation
between posttranslational and translational processes in the model~\cite{ao03},
we are able to analytically and exactly compute how frequently the model will predict the
correct gene pattern.

\section{The segment polarity gene network in {\it Drosophila}}
\label{sec-drosophila}
We will apply our analysis to a Boolean model of the interactions among the
{\it Drosophila melanogaster} segment polarity genes. 
This gene network represents the last step in the 
hierarchical cascade of gene families initiating the segmented body of the fruit fly. 
While genes in preceding stages of development act transiently, the segment polarity genes 
are expressed throughout the life of the fly. 
The best characterized segment polarity genes include {\it engrailed} ($\en$), 
{\it wingless} ($\wg$), {\it hedgehog} ($\hh$), {\it patched} ($\ptc$), 
{\it cubitus interruptus} ($\ci$) and {\it sloppy paired} ($slp$), coding for  
the corresponding proteins, which we will represent by capital letters 
($\EN$, $\WG$, $\HH$, $\PTC$, $\CI$ and $\SLP$). 
Two additional proteins, resulting from transformations of the protein $\CI$, 
also play important roles: $\CI$ may be converted into a transcriptional activator, $\CIA$,
or may be cleaved to form a transcriptional repressor $\CIR$.

The expression pattern of the {\it Drosophila} segment polarity genes 
(see Table~\ref{table_steady_states})
is maintained almost unmodified for three hours, 
during which time the embryo is divided into $14$ parasegments. 
Each of these parasegments is composed of about 4 cells, delimited by furrows positioned 
between the $\wg$ and $\en$ -expressing cells~\cite{hs92}. 

The Boolean model that we will study was introduced and developed by one of us in~\cite{ao03}. 
(Further robustness analysis was also developed in~\cite{cas05}.) 
In this model, a parasegment of four cells is considered: the variables are the expression levels 
of the segment polarity genes and proteins (listed above) in each of the four cells
(the total number of nodes in the network is thus $4\times 13=52$.).
The expression level of each gene or protein is assumed to be either 0 (OFF) or 1 (ON).
The model successfully describes the transition from the initial expression 
pattern\rf{eq-initial-wt} to a final pattern two or three developmental stages later, 
when the embryo has been clearly divided into parasegments 
(see first entry of Table~\ref{table_steady_states}).
As discussed in~\cite{ao03}, the evolution of these gene expression patterns is well described 
by a set of logical rules, which are depicted in Table~\ref{table_rules}.

We adopt the notation ``$\wg_1^k$'' or ``$\wg_1(k)$'' to represent the state of 
{\it wingless} mRNA in the first cell of the parasegment at time $k$. 
Similar notations apply for other mRNAs and proteins.
Periodic boundary conditions are assumed, meaning that: $node_{4+1}=node_1$ and 
$node_{1-1}=node_4$. The wild type initial pattern corresponds to: 
\beqn{eq-initial-wt}
   \wg_4^0=1,\  \en_1^0=1,\  \hh_1^0=1, \ \ptc_{2,3,4}^0=1, \ \ci_{2,3,4}^0=1,
\eeqn
with the remaining nodes zero.

\begin{table}
\caption{Regulatory functions governing the states of segment polarity gene products in the model. 
Each node is labeled by its biochemical symbol and subscripts signify cell number.}
\label{table_rules}
\centering{
\begin{tabular}{ll}
\hline %\toprule
Node & Boolean updating function (synchronous algorithm) \\
\hline\hline%\midrule
$\SLP_i$ & $\SLP_i(k+1)=\left\{\begin{array}{lllll}
0 &\mbox{if}& i\in\{1,2\}\\
1 &\mbox{if}& i\in \{3,4\}\\
\end{array}\right.$ \\
$\wg_i$ & $\wg_i(k+1)=(\CIA_i(k)$ and $\SLP_i(k)$ 
       and not $\CIR_i(k))$ \\ 
       & or $[\wg_i(k) $ and $(\CIA_i(k)$ or $\SLP_i(k) )$ 
		 and not $\CIR_i(k)]$ \\
$\WG_i$ & $\WG_i(k+1)=\wg_i(k)$  \\
$\en_i$ & $\en_i(k+1)=(\WG_{i-1}(k)$ or $\WG_{i+1}(k))$ 
         and not $\SLP_i(k)$ \\
$\EN_i$ & $\EN_i(k+1)=\en_i(k)$  \\
$\hh_i$ & $\hh_i(k+1)=\EN_i(k)$ and not $\CIR_i(k)$ \\
$\HH_i$ & $\HH_i(k+1)=\hh_i(k)$ \\
$\ptc_i$ & $\ptc_i(k+1)=\CIA_i(k)$ and  not $\EN_i(k)$ and not $\CIR_i(k)$ \\
$\PTC_i$ & $\PTC_i(k+1)=\ptc_i(k)$ or $(\PTC_i(k)$  and not $\HH_{i-1}(k)$ \\
        & and not $\HH_{i+1}(k))$ \\
$\ci_i$ & $\ci_i(k+1)=$ not $\EN_i(k)$ \\
$\CI_i$ & $\CI_i(k+1)=ci_i(k)$ \\
$\CIA_i$ & $\CIA_i(k+1)=CI_i(k)$ and [not $\PTC_i(k)$ or $\HH_{i-1}(k)$\\ 
        &   or $\HH_{i+1}(k)$ or $\hh_{i-1}(k)$ or $\hh_{i+1}(k)$]\\
$\CIR_i$ &  $\CIR_i(k+1)=\CI_i(k)$ and $\PTC_i(k)$ and not $\HH_{i-1}(k)$ \\
        & and not $\HH_{i+1}(k)$ and not $\hh_{i-1}(k)$ and not $\hh_{i+1}(k)$\\
\hline%\bottomrule 
\end{tabular}}
\end{table}

\subsection{Steady states of the Boolean model}
A complete analysis of the steady states is found in~\cite{ao03}. 
Table~\ref{table_steady_states} summarizes these results, indicating the
expressed nodes in each of the six steady-states.
We note that three of the four main steady states agree perfectly with 
experimentally observed states corresponding to wild type, to $\ptc$ knockout mutant (broad striped) 
and to $\en$, $\wg$ or $\hh$ knockout mutant (non-segmented)
embryonic patterns (\cite{gakt00,tek92}; see~\cite{ao03} for more references).
\begin{table}[h]
\caption{Complete characterization of the model's steady states.}
\label{table_steady_states}
\begin{center}
\begin{tabular}{ll}
\hline
  Steady state  & Expressed nodes \\
\hline\hline
  wild type & $\wg_{4}$, $\WG_{4}$, $\en_{1}$, $\EN_{1}$, $\hh_{1}$, $\HH_{1}$, \\
            & $\ptc_{2,4}$, $\PTC_{2,3,4}$, $\ci_{2,3,4}$, \\
            & $\CI_{2,3,4}$, $\CIA_{2,4}$, $\CIR_{3}$   \\
\hline
  broad stripes & $\wg_{3,4}$, $\WG_{3,4}$, $\en_{1,2}$, $\EN_{1,2}$, 
                       $\hh_{1,2}$, $\HH_{1,2}$,  \\
                & $\ptc_{3,4}$, $\PTC_{3,4}$, $\ci_{3,4}$, $\CI_{3,4}$, $\CIA_{3,4}$  \\
\hline
  no segmentation & $\ci_{1,2,3,4}$, $\CI_{1,2,3,4}$, 
                    $\PTC_{1,2,3,4}$, $\CIR_{1,2,3,4}$ \\
\hline
  wild type variant & $\wg_{4}$, $\WG_{4}$, $\en_{1}$, $\EN_{1}$, $\hh_{1}$, $\HH_{1}$, \\
            & $\ptc_{2,4}$, $\PTC_{1,2,3,4}$, $\ci_{2,3,4}$, \\
            & $\CI_{2,3,4}$, $\CIA_{2,4}$, $\CIR_{3}$ \\
\hline
  ectopic  &  $\wg_{3}$, $\WG_{3}$, $\en_{2}$, $\EN_{2}$, $\hh_{2}$, $\HH_{2}$, \\
           &  $\ptc_{1,3}$, $\PTC_{1,3,4}$, $\ci_{1,3,4}$, \\
           &   $\CI_{1,3,4}$, $\CIA_{1,3}$, $\CIR_{4}$   \\
\hline
  ectopic variant & $\wg_{3}$, $\WG_{3}$, $\en_{2}$, $\EN_{2}$, $\hh_{2}$, $\HH_{2}$, \\
            & $\ptc_{1,3}$, $\PTC_{1,2,3,4}$, $\ci_{1,3,4}$, \\
            &  $\CI_{1,3,4}$, $\CIA_{1,3}$, $\CIR_{4}$  \\
\hline
\end{tabular}
\end{center}
\end{table}

\subsection{The regulatory function of the {\it sloppy paired} gene}

The rule for $\SLP$ protein in Table~\ref{table_rules} summarizes in a simple way the experimental observations 
on the expression and regulatory activity of the {\it sloppy paired} gene in the segment polarity network~\cite{cgg94}.
A more detailed rule for the {\it sloppy paired} expression pattern can be created to incorporate recent evidence of engrailed protein inhibiting $\slp$ transcription~\cite{av03}. 
However, inhibition by engrailed accounts only partially for the experimentally observed restriction of 
$\slp$ to the posterior half of the parasegment. Thus we need to invoke an additional regulatory 
effect, which we denote by $\RX$. $\RX$ probably represents a combination of regulation by the 
pair-rules responsible for the establishment of $\slp$, namely {\it runt}, {\it opa} and 
Factor X~\cite{sg04} and of $\slp$ autoregulation.  

Therefore, $\SLP$ expression in Table~\ref{table_rules} can be replaced by
the following set of equations:
\beqn{eq-RX}
  \RX_i(k+1)&=&\left\{\begin{array}{lllll}
                 0, &\mbox{ if }& i\in\{1,2\}\\
                 1, &\mbox{ if }& i\in \{3,4\}\\
               \end{array}\right. \nonumber \\ & & \\
 \slp_i(k+1)&=&\RX_i(k)  \mbox{ and not }\EN_i(k) \nonumber\\
 \SLP_i(k+1)&=&\slp_i(k) \ . \nonumber
\eeqn
The {\it sloppy paired} initial conditions would then be:
\beqn{eq-initial-slp}
  \slp_{3,4}^0=1,\ \SLP_{3,4}^0=1. 
\eeqn

This generalization of the segment polarity network model introduces additional steady states,
such as a two-stripe $\en$ pattern characterized by $\slp_4=\SLP_4=1$, $\wg_4=\WG_4=1$, 
$\en_{1,3}=\EN_{1,3}=1$, $\hh_{1,3}=\HH_{1,3}=1$, $\ptc_{2,4}=\PTC_{2,4}=1$,
and $\ci_{2,4}=\CI_{2,4}=\CIA_{2,4}=1$ 
(this pattern was also found in~\cite{ao03} as a result of $\slp$ mutation). This expression pattern is non-viable since it has two $\it en- wg$ borders and would lead to an ectopic parasegment structure. On the other hand, it is not difficult to see that, starting from conditions\rf{eq-initial-wt} 
and\rf{eq-initial-slp}, none of the ``new'' steady states are reachable, since these
initial conditions imply that neither $\slp$ nor $\SLP$ can change their expression at any time.
Indeed we have the following result:

\begin{lem}{lm-slp}
Consider the extended model of Table~\ref{table_rules} together with\rf{eq-RX}.
Assume that initial conditions are given by\rf{eq-initial-wt} and\rf{eq-initial-slp}.
Then $\slp_i(t)=\slp_i^0$ and $\SLP_i(t)=\SLP_i^0$ for all times . 
\end{lem}

A sketch of proof is as follows. Note first that $\slp_{1,2}(t)=\SLP_{1,2}(t)\equiv0$ follows 
from $\RX_{1,2}(t)=0$ for all $t$.
Next, observe that:
\beq 
  \slp_4(t_1)=0 \Rightarrow \EN_4(t_2)=1 \Rightarrow \en_4(t_3)=1 
   \Rightarrow \SLP_4(t_4)=0 \Rightarrow \slp_4(t_5)=0, 
\eeq
where $t_1\geq t_2\geq t_3\geq t_4\geq t_5\geq0$.
Thus, in order for $\slp_4$ to become zero at any time, it had to be so at 
some previous instant ($t_5\leq t_1$). If $\slp_4(0)=1$, and if $T>0$ is the first
instant such that $\slp_4(T)=0 $ then we have a contradiction. Hence $\slp_4(t)=1$
for all times. (Note that this argument is independent of the order in
which the nodes are updated.)
Similar arguments show that $\slp_3(t)=\slp_3(0)=1$ and $\SLP_{3,4}(t)=1$,
for all $t$. 

Thus the extended model leads to the same results as assuming a constant $SLP$ pattern. For this reason, and lacking more specific biological evidence on the regulation of {\it sloppy paired}, our present analysis is focused on the simpler biologically relevant
model of Table~\ref{table_rules}.

\section{Asynchronous algorithms}
\label{sec-asynch}
In general, for a network of $N$ gene products (denoted $x_1$, $\ldots$, $x_N$), 
the dynamics of a Boolean model is typically studied by simultaneously updating
the state of all the nodes in the network, according to 
\beq
   X_i^{k+1}=F_i(X_1^k,X_2^k,\ldots,X_N^k),\ \ \ i=1,\ldots,N
\eeq
where $F_i$ is the regulating function for mRNA or protein $X_i$.
An underlying hypothesis is the existence of perfect synchronization among 
the various regulatory processes. However, it is well known that 
the timescales of transcription, translation, and degradation processes
can vary widely from gene to gene and can be anywhere from minutes to hours.

In analogy with task coordination and data communication procedures 
in the context of parallel computation systems~\cite{tsi}, we have previously
developed several methods that introduce different timescales for the different 
regulatory processes within the network~\cite{cas05}. 
%A first algorithm consists of  allowing the (constant) time unit of the 
%synchronous model to be randomly perturbed for each node.
These include algorithms that randomly choose the order in which the nodes are updated
(this random order algorithm is summarized below in~\ref{sec-rand0}, 
and in Section~\ref{sec-timesep}), 
and a totally asynchronous algorithm, where the next updating times for each node are 
randomly chosen at each instant.

We now introduce a more intuitive asynchronous algorithm, where each node is updated 
according to its own specific time unit.
The time units for the nodes are randomly chosen from a uniform distribution in an 
interval $[1-\eps,1+\eps]$, where $\eps\in(0,1)$. The updating times of $i$-th node are
then pre-specified as:
\beq
    T_i^1, T_i^2,\ldots, T_i^k,\ldots \ \ \  k\in\N,
\eeq
with
\beqn{eq-itimes}
   T_i^{k+1} = T_i^k +\gamma_i = k\gamma_i,\ \ \  k\in\N.
\eeqn
For instance, $\gamma_{\tWG_4}<\gamma_{\twg_4}$ means that wingless protein in the fourth cell
is translated at a faster rate (shorter time intervals) then {\it wingless} mRNA is produced.
At any given time $t$, the next node(s) to be updated is(are) $i$ 
such that $T_i^k=\min_{j,\ell}\{ T_j^\ell\geq t\}$, for some $k$.
The variables $X_i$ are updated according to:
\beqn{eq-inode}
   X_i(T_i^k) = F_i(\, X_1(\tau^k_{1i}),\ldots,X_N(\tau^k_{Ni})\,),
\eeqn
where $\tau^k_{ji}$ defines the most recent instant when node $j$ was updated, that is
\beqn{eq-ijtau}
   \tau^k_{ji}= \max_\ell\ \{ \ T_j^\ell: \ T_j^\ell < T_i^k \ \}.
\eeqn
By ordering all the time sequences $\{T^k_{i}:\ i=1\ldots N,\ k=1,2,\ldots\}$, into a single 
nondecreasing sequence, say $\{t_1,t_2,\ldots\}$, the asynchronous model can also 
be written in the form
\beq
   X_i(t_{k+1})=\left\{
                \begin{array}{ll}
                   F_i(X_1(t_k),X_2(t_k),\ldots,X_N(t_k)), 
                              & \mbox{if } t_k=\ell\gamma_i \mbox{ for some $i$, $\ell$} \\
                   X_i(t_k), & \mbox{otherwise.}
                \end{array}
                   \right.
\eeq
It is clear that the steady states of this model must satisfy 
$X_i(t_{k+1})=X_i(t_k)=F_i(X(t_k))$,
and therefore are the same as those of the synchronous boolean model in 
Table~\ref{table_steady_states}.

Note that the case $\eps=0$ reduces to the synchronous model, where every node is updated
simultaneously ($\gamma_i=1$), at the same time instants: $T_i^k=k$, for all $i=1,\ldots,N$.
This algorithm allows great variability in each process' duration, exploring the gene 
expression patterns due to all possible combinations of individual timescales. 
{Implementation of this asynchronous algorithm shows that, if started from the initial wild 
type state\rf{eq-initial-wt}, any of the steady states of the model 
(Table~\ref{table_steady_states}) may occur with a certain probability.}
{The probability of occurrence of each pattern depends on the range over which
the individual time units $\gamma_i$ are allowed to vary  (see Fig.~\ref{fig-interval-length}). 
For $\eps=0$, the wild type steady state is attained with probability 100\% 
(corresponding to the synchronous 
Boolean model). As $\eps$ increases to 0.01 (resp. 0.1) this value decreases to 60\% (resp. 44\%).
However, further increase in $\eps$ (hence larger time intervals) unexpectedly leads to an 
increase in the occurrence of the wild type state, up to 51\% for $\eps=0.9$.}
Other final states observed are the 
broad-striped pattern ($25\%-38\%$) observed in heat-shock experiments and $\ptc$ 
mutants~\cite{gakt00} and the pattern with no segmentation ($12\%-15\%$) observed 
in $\en$, $\hh$ or $\wg$ mutants~\cite{tek92}, the latter two corresponding to 
embryonic lethal phenotypes~\cite{gakt00}. 
Each of the other three steady states occurs with frequencies less than $5\%$.
(These values were obtained from 10000 numerical experiments.)

A possible extension of this algorithm would be to consider a discrete model with a finite 
number of logical levels describing ON, OFF as well as other intermediate steps of the 
system~\cite{st01}. This would involve decisions about the number of intermediate steps,
their values, and development of new transition rules.
Instead, in this paper we will focus on a ``hybrid'' model, that takes into account 
the continuous nature of the biological processes, while still using Boolean rules to
describe ON/OFF transitions (Section~\ref{sec-glass}).

\subsection{Random order algorithm}
\label{sec-rand0}
For comparison purposes, we briefly summarize an alternative asynchronous algorithm, 
which guarantees that every node is updated exactly once during each unit time 
interval~\cite{cas05}.
A random order of updates for the $N$ nodes is generated as a permutation
$\phi^k$ of $\{1,\ldots N\}$. This permutation is randomly chosen out of a uniform 
distribution over the set of all $N!$ possible permutations, at the beginning of 
the time unit $k$. The updating times for each node are now written as
\beq
   T_i^k = N(k-1) + \phi^k(i),\ \ \  k\in\N,
\eeq
so that $\phi^k(j)<\phi^k(i)$ implies $T_j^k<T_i^k$, and node $j$ is updated before node $i$
at the $k$-th iteration.
The results of this algorithm are qualitatively similar to those of the asynchronous 
algorithm (Table~\ref{table_steady_states}).

\begin{table}
\caption{The frequencies of the six steady states observed with the three different methods when
starting from the wild type initial condition.}
\label{tab-all-permutations}
\begin{center}
\begin{tabular}{lccc}
\hline
 Steady state      & Asynchronous & Random order & Glass-type\\
   pattern         & algorithm   & algorithm &      model\\
\hline\hline
  wild type      & 44-51\%  & 56\%  &  89-100\% \\
%\hline
  broad stripes   & 25-38\% &  24\%  & 0-6\% \\
%\hline
  no segmentation  & 12-15\%  & 15\%   & 0-3\% \\
%\hline
  wild type variant &  4-5.6\% & 4.2\%  & 0-1\% \\
%\hline
  ectopic     &  0.4-1\% & 0.98\%  & 0\% \\
%\hline
  ectopic variant & 0.1-0.5\%  & 0.68\%  & 0\% \\
\hline 
\end{tabular}
\end{center}
\end{table} 

\section{Glass-type networks}
\label{sec-glass}
\begin{figure}
\centerline{
\psfig{figure=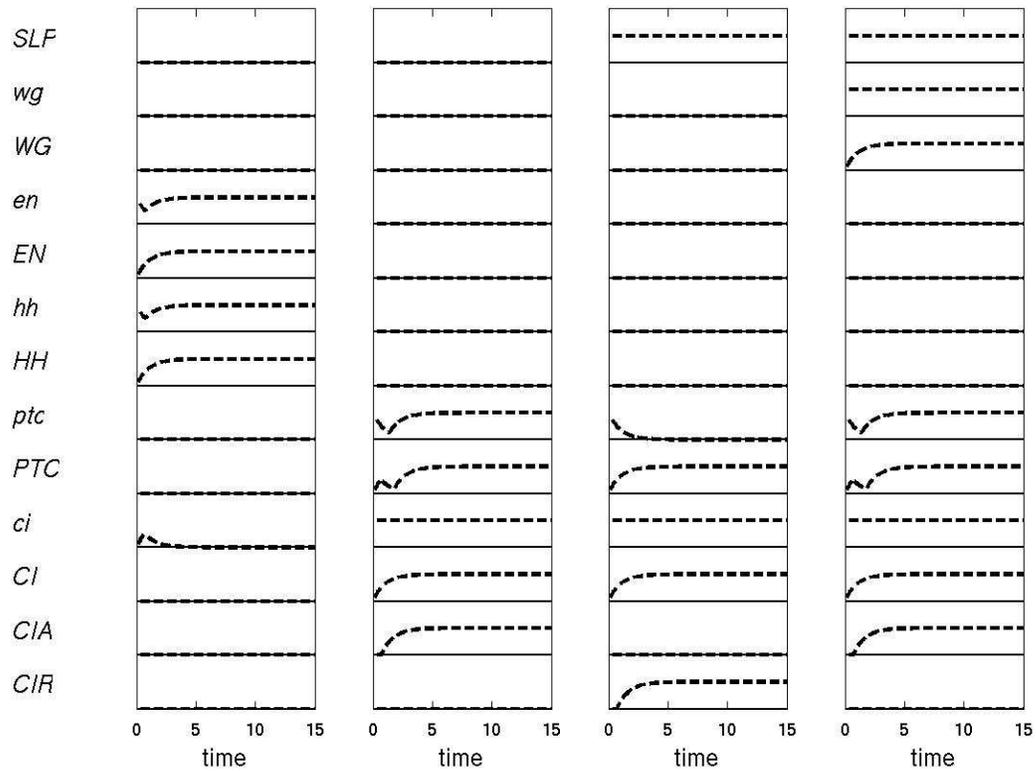,height=4.5in}}  
\caption{The solution to the system of piecewise linear equations\rf{eq-glass} (dashed lines). 
Each column represents one cell, and each of the 13 rectangles represents the continuous-time 
variable for proteins or mRNAs (as labeled at left). In each rectangle, the y-axis ranges 
from 0 to 2 units.
The nodes for which the trajectories converge to 1 (middle of the rectangle) are exactly 
those expressed in the wild type steady state. (The time units are arbitrary.)}
\label{fig-glass-wt}
\end{figure}

The asynchronous algorithm defined by\rf{eq-itimes},\rf{eq-inode} and\rf{eq-ijtau}
allows the introduction of distinct timescales for each regulation process in a 
Boolean model.
We next propose an alternative method which provides a bridge between discrete and continuous 
approaches, resulting in a more realistic model, but without the necessity of specifying
any kinetic or binding parameters (which are typically unknown).
In this  method, the gene and protein levels are represented 
as continuous variables, and their time evolution is described by differential equations, 
but the interactions among nodes are still modeled by Boolean functions~\cite{g75,eg00,gk73,jghpsg04}.
Glass~\cite{g75} introduced a class of piecewise linear differential equations that
combine logical rules for the synthesis of gene products with linear (free) decay by describing
each node with two variables, one discrete and one continuous.
For simplicity of notation, in what follows we will let $\cX_i$ denote the continuous 
variable associated with node $i$, its discrete variable $X_i$, and the discrete variable's
Boolean rule by $F_i$. The Glass-type model is then
\beqn{eq-glass}
   \frac{d\, \cX_i}{dt} = -\cX_i + F_i(X_1,X_2,\ldots,X_N),\ \ \ i=1,\ldots,N.
\eeqn 
At each instant $t$, the discrete variable $X_i$ is defined as a function of
the continuous variable according to a threshold value:
\beqn{eq-threshold}
   X_i(t)=\left\{ \begin{array}{ll}
                          0, & \cX_i(t)\leq\theta \\
                          1, & \cX_i(t)>\theta\ ,
               \end{array}
       \right.                   
\eeqn
where $\theta\in(0,1)$.
The discrete variables $X_i$ represent the ON and OFF levels of the nodes 
in the Boolean model.
The underlying assumption in this Glass model is that the decay rate and activation 
threshold of each gene product is identical.
Since the initial condition for the piecewise linear system\rf{eq-glass} is also\rf{eq-initial-wt}
(i.e., $X(0)=\cX(0)$) and $F_i\in\{0,1\}$, it is easy to see that solutions of\rf{eq-glass}
evolve in the hypercube $[0,1]^N$.
Under these conditions, the limiting values ``0'' and ``1'' of the continuous variable $\cX_i$
represent, respectively, ``absence of species $i$'' and ``maximal concentration of species $i$''
-- thus we can view the $\cX_i$ as dimensionless variables, scaled to attain their maximal values at 1.
The continuous dynamics is translated into a Boolean ON/OFF response, according to $\theta$:
as soon as $\cX_i$ increases above $\theta$, species $i$ is considered to be in the ON state; otherwise
it remains in the OFF state (see also~\cite{jghpsg04}). 
Thus the parameter $\theta$ defines the fraction of ``maximal  concentration'' 
necessary for a protein or mRNA to effectively perform its biological function.
This method allows us to study the continuous evolution of the genetic network
simply by specifying $\theta$, the fraction of maximal concentration that is effective as ON level, 
avoiding the need to specify any kinetic parameters.
Below and in Section~\ref{sec-convergence}, we will see that system\rf{eq-glass} exhibits 
distinct dynamics in the two regions $\theta\leq1/2$ and $\theta>1/2$. 
It is easy to see that the steady states of the piecewise linear equations\rf{eq-glass}
are still those of the Boolean model, since:
\beq
   \frac{d\, \cX_i}{dt} = 0 \ \Leftrightarrow \ \cX_i=F_i(X_1,X_2,\ldots,X_N),\ i=1,\ldots,N,
\eeq 
independently of $\theta$. 
Applying this method to the {\it Drosophila} segment polarity gene network, we find an
exact convergence to the wild type steady state when started from 
the wild type initial condition (Fig.~\ref{fig-glass-wt}),
independently of the ON/OFF threshold value $\theta$.
This result supports the Boolean model as a suitable description of the underlying network of
gene interactions.

\subsection{Introducing distinct timescales}
The assumption of equal decay rates and activation thresholds for all nodes is
an oversimplification similar to that made in synchronous Boolean models.
However, for further robustness analysis, one may introduce different timescales for
the different processes, by scaling the time units in each differential equation according to: 
\beqn{eq-glass-a}
   \frac{d\, \cX_i}{dt} = \alpha_i(\ -\cX_i + F_i(X_1,X_2,\ldots,X_N) \ ), 
\eeqn 
with $\alpha_i\geq\eps$ for some fixed $\eps>0$, $i=1,\ldots,N$. 
Each $X_i$ is a discrete variable defined as before
(note that steady states of this new system are still those of the
Boolean model).\footnote{We will analyze the behaviors of trajectories of systems of the
form\rf{eq-glass}, assuming that trajectories are well-defined.  Since the
right-hand sides of equations of these type are discontinuous, it is very
difficult to give general existence and uniqueness theorems for solutions of
inital-value problems.  One must impose additional assumptions, insuring that
only a finite number of switches can take place on any finite time interval,
and often tools from the theory of differential inclusions must be applied, 
see for instance~\cite{gedeon} for more discussion. See also~\cite{jghpsg04}.}

\begin{figure}
\centerline{
\psfig{figure=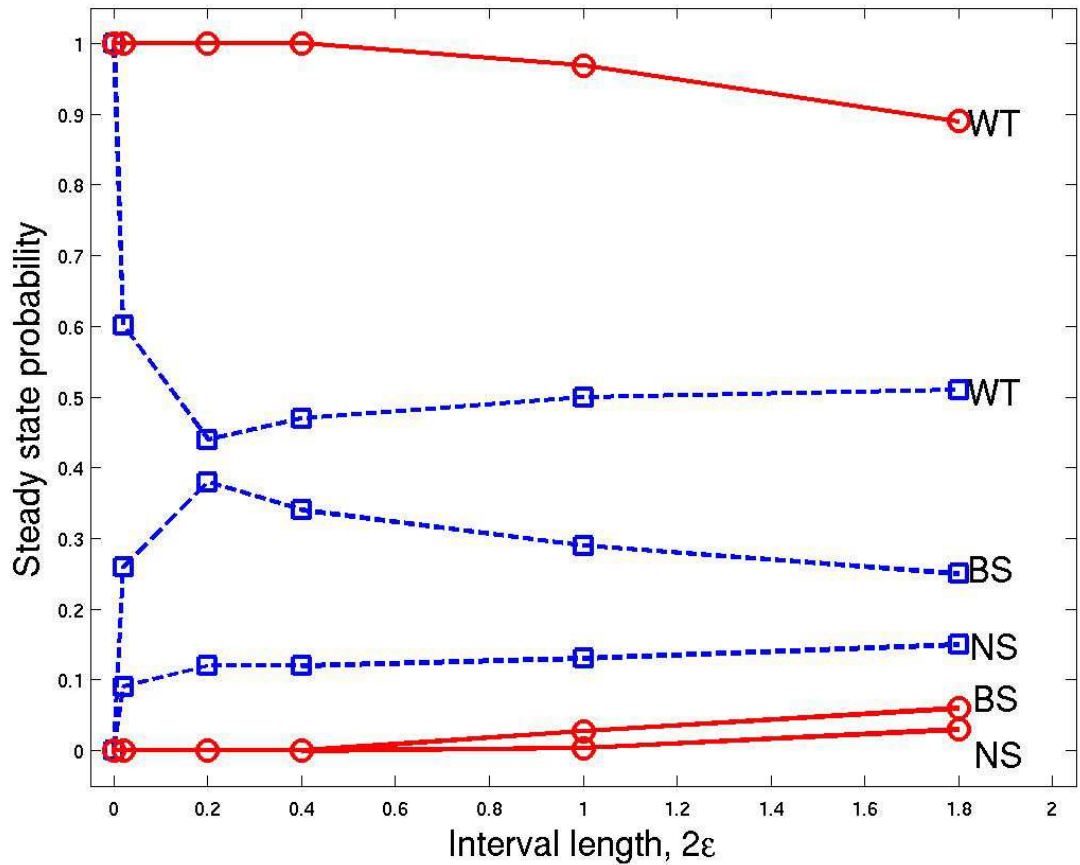,height=5in} } 
\caption{Probability of occurrence of the three most frequent patterns:
wild type (WT), broad stripes (BS), and no segmentation (NS), under variable range of timescales. 
Dashed lines/squares represent asynchronous algorithm results, while solid lines/circles 
represent Glass-type model results (out of 1000 runs). Results were obtained with $\theta=0.5$.}
\label{fig-interval-length}
\end{figure}

\begin{figure}
\centerline{
\psfig{figure=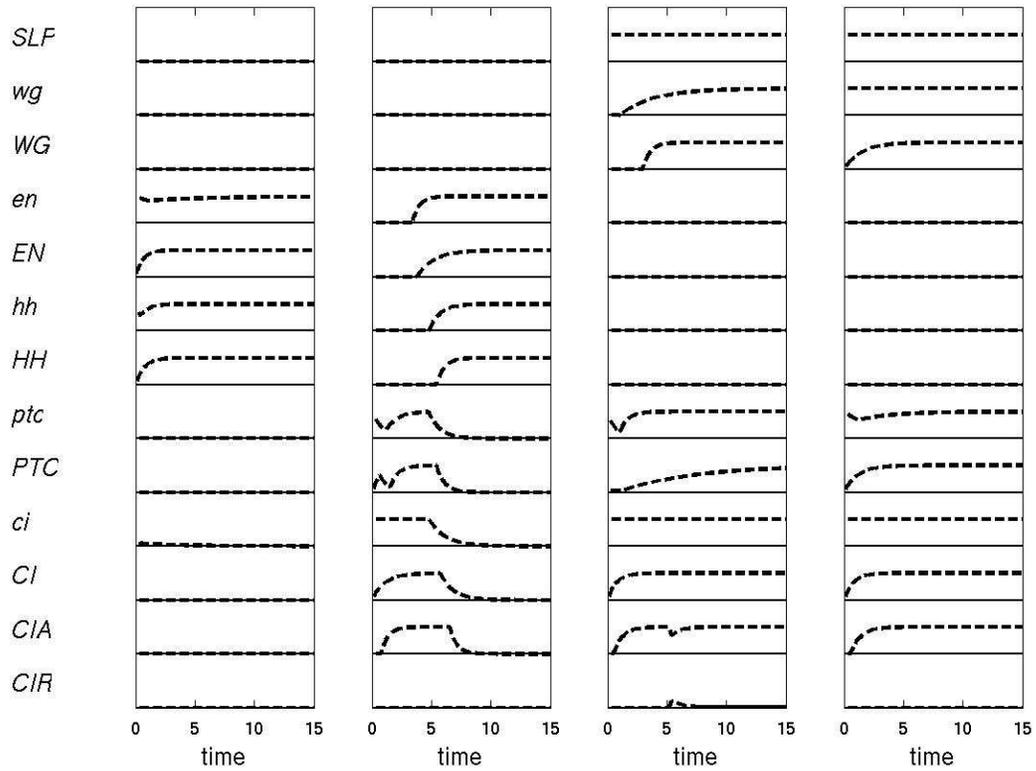,height=4.5in} }
\caption{A solution to the system of piecewise linear equations\rf{eq-glass-a}
(dashed lines), in an example where the steady state corresponds to the broad stripes pattern.
In each rectangle, the y-axis ranges from 0 to 2 units.
Notice that {\it wingless} is expressed in two adjacent cells, as opposed to the wild type
pattern, where {\it wingless} is expressed in only one cell (similarly for {\it engrailed} 
and {\it hedgehog}).}
\label{fig-glass-bs}
\end{figure}

\begin{figure}
\centerline{
\psfig{figure=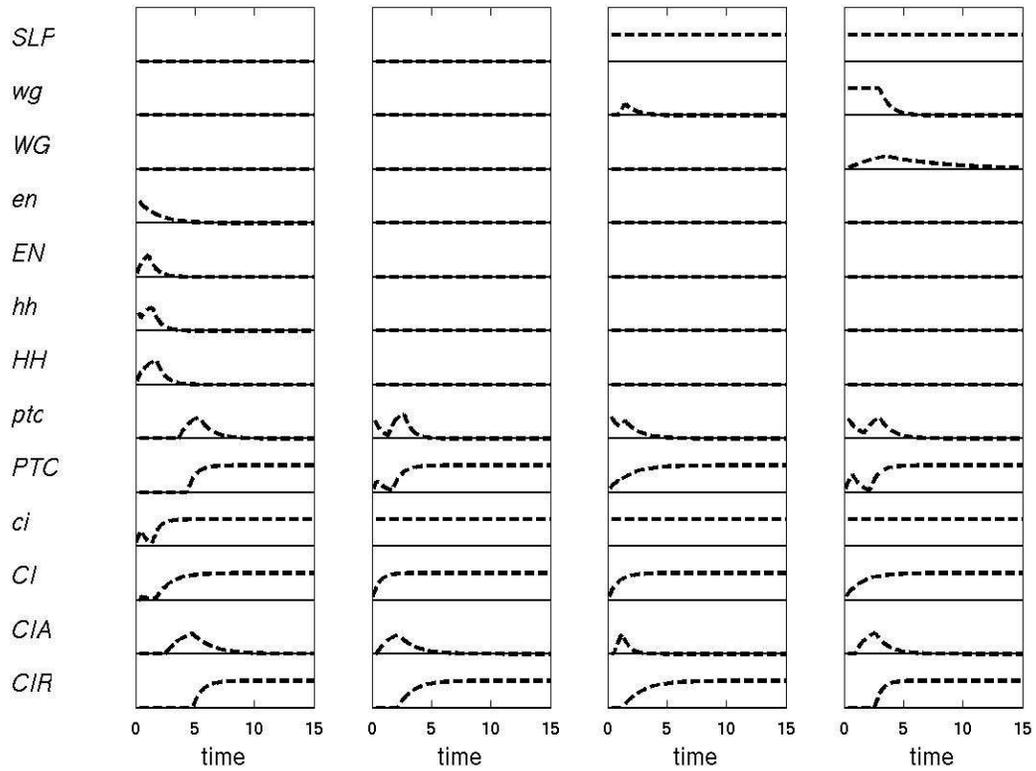,height=4.5in} }
\caption{A solution to the system of piecewise linear equations\rf{eq-glass-a}
(dashed lines), in an example where the steady state corresponds to the no segmentation pattern.
In each rectangle, the y-axis ranges from 0 to 2 units.
Notice that {\it wingless}, {\it engrailed} and {\it hedgehog} are not expressed in any cell, 
thus no segments are visibly detected in the embryo.}
\label{fig-glass-ns}
\end{figure}

This method represents a continuous equivalent to the asynchronous algorithm described in 
Section~\ref{sec-asynch}. Here, the (inverse) scaling factors $\alpha_i^{-1}$ may be viewed 
as half-lives of mRNA or proteins. These may be directly compared to the individual time
units $\gamma_i$ as follows. Using Euler's method to discretize system\rf{eq-glass-a} obtains:
\beq
   \cX_i(t+\Delta t)= \cX_i(t)+\alpha_i\Delta t (- \cX_i(t)+F_i(X(t))).
\eeq
Now notice that choosing the integrating time interval to be such that $\alpha_i\Delta t=1$
recovers the discrete asynchronous algorithm with specific time units
\beq
   \gamma_i=\Delta t=\alpha_i^{-1}.
\eeq
For comparison to the discrete algorithm, we choose both the scale factors $\alpha_i^{-1}$
and the time units $\gamma_i$ 
randomly from a uniform distribution in intervals of the form $[1-\eps,1+\eps]$, $\eps\in(0,1)$. 
The numerical experiments are reported in Fig.~\ref{fig-interval-length}, where the threshold 
value $\theta$\rf{eq-threshold} was set to $1/2$.
We again observed that, starting from wild type initial
  conditions\rf{eq-initial-wt}, all steady states may occur with a certain frequency 
(see Table~\ref{tab-all-permutations}). But, in contrast to the asynchronous Boolean model, 
the wild type pattern occurs with frequencies that decrease monotonically with $\eps$,
down to 89\% for $\eps=0.9$ (Fig.~\ref{fig-interval-length}). 
The next more frequently achieved patterns are the broad stripes
(Fig.~\ref{fig-glass-bs}), with probability 6\% for $\eps=0.9$, the no segmentation 
(Fig.~\ref{fig-glass-ns}), with probability 3\%, and the wild type
variant, with probability 1\%.

The three methods we have described (Table~\ref{tab-all-permutations})
produce qualitatively compatible results, in the sense that the wild
type pattern is always the most frequently occurring steady state,
followed by the broad stripes, no segmentation, and wild type variant patterns.

\subsection{Fraction of maximal concentration that defines an ON state}
\begin{figure}
\centerline{
\psfig{figure=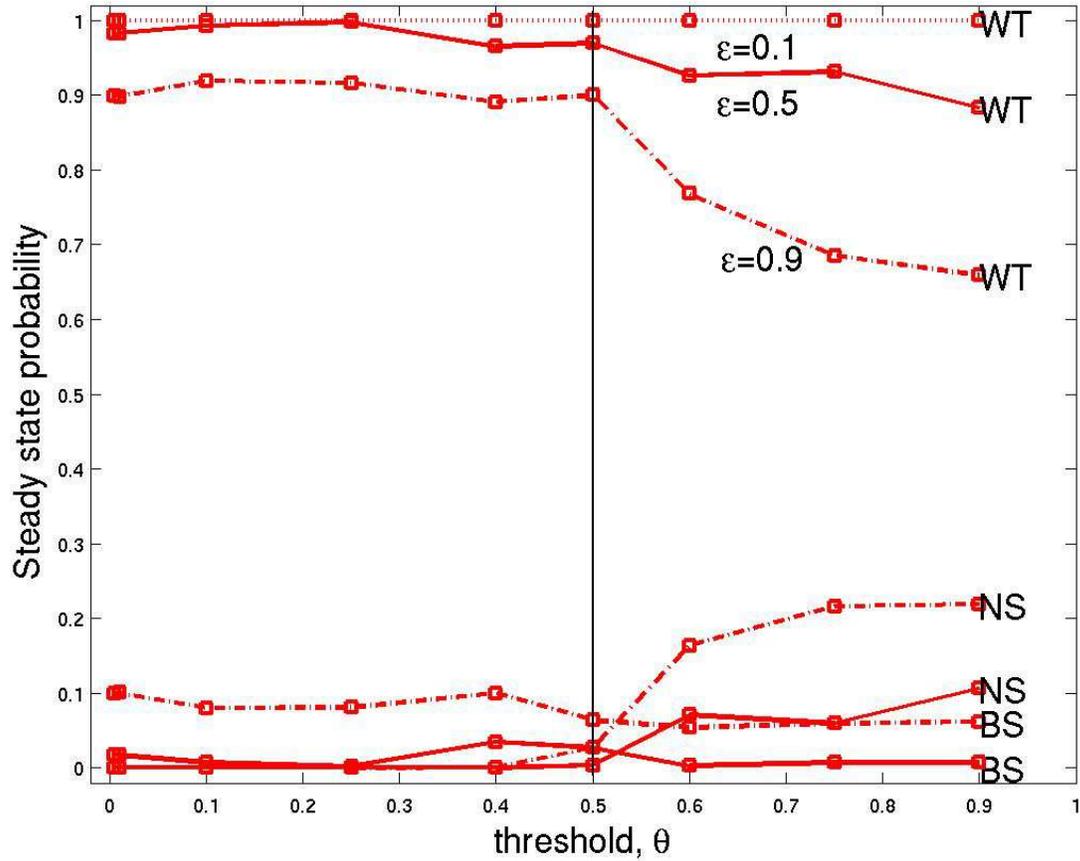,height=5in} }
\caption{Effect of effective ON concentration, $\theta$,
on the probability of occurrence of the three most frequent patterns: 
wild type (WT), broad stripes (BS), and no segmentation (NS).
The results are for the Glass-type model with $\alpha_i^{-1}$ randomly chosen
in an interval $[1-\eps,1+\eps]$.
Solid lines represent the case $\eps=0.5$, dotted lines represent the case $\eps=0.1$
and dash-dotted lines represent the case $\eps=0.9$.
(results out of 1000 runs).}
\label{fig-threshold}
\end{figure}

The piecewise linear system\rf{eq-glass-a} follows the threshold\rf{eq-threshold} to decide 
whether a given node is ON or OFF. While this value $\theta$ did not affect the 
dynamics of the system in the case $\alpha_1=\alpha_2=\cdots=\alpha_N$, it plays a significant role 
in the general case.
In Figure~\ref{fig-threshold}, it is immediate to see that the effect of $\theta$ depends on the 
length of the interval allowed for the timescales. Indeed, there is a marked difference between
narrow ($\eps\leq0.5$) and wide ($\eps\geq0.6$) intervals.
For narrow intervals, numerical experiments indicate that, starting from initial condition\rf{eq-initial-wt}, 
system\rf{eq-glass-a} converges  to wild type steady state with probability around 90\% or more;
whereas, for wider intervals, this probability may decrease down to 68\% ($\eps=0.9$, $\theta=0.9$).

On the other hand, the threshold value also divides the dynamics into two regions.
For $\theta\leq0.5$, the probability that initial condition\rf{eq-initial-wt} leads to
the wild type steady state is above 90\%, independently of $\eps$. For $\theta\leq0.5$ 
it is less probable to reach the non segmented (around 1\%) than the broad striped pattern 
(around 10\%).
For higher $\theta\geq0.6$, the probability of reaching the wild type steady state (from 
initial condition\rf{eq-initial-wt}) decreases very significantly, and in addition, it becomes 
more problable to reach the non segmented (around 20\%) than the broad striped pattern (around 7\%).

In our Glass-type model, $\theta$ represents the fraction of maximal concentration above which 
an mRNA or protein is considered ON, or biologically effective. 
Our results indicate that quite small fractions of the maximal concentration can (and should) be 
interpreted as sufficient amounts for an mRNA or protein to be in the ON state.
Namely, when the concentration of mRNAs or proteins has increased to a fraction up to half 
its maximum possible value, it is already present in a sufficient amount to perform its function.
This follows from observation of Figure~\ref{fig-threshold}: $0<\theta\leq0.5$ leads to
a good (realistic) performance of the model, with 90\% convergence to wild type pattern.
Our results also indicate that a higher threshold is, on the contrary, not very realistic. 
For instance, setting $\theta\geq0.6$ leads to a fairly high incidence on the mutant patterns.
But by letting $\theta\geq0.6$ one is assuming that an mRNA or protein is not present in a sufficient 
amount to perform its function until it reaches at least sixty per cent of its maximal value.
Typically such high thresholds are not observed: indeed, in~\cite{dmmo00} (supplementary 
material) where continuous dynamics are also transformed into an ON/OFF response, 
a threshold of 10\% is used, and considered very reasonable.
In our experiments, unless otherwise indicated, we have used $\theta=0.5$.

To further generalize system\rf{eq-glass-a}, it is  natural to allow each variable to respond to its
own threshold and consider distinct $\theta_i$ values. 
We will address this problem in Section~\ref{sec-theta-vary}, and see that the system's dynamics is 
preserved in each $\theta$ region.
In fact, our simulations with distinct $\theta_i$ recover many of the theoretical results we obtained 
for a species-independent $\theta$ (Section~\ref{sec-convergence}).

\section{Pre-patterning errors and knockout mutant situations}
\label{sec-mutants}

In~\cite{ao03,cas05} we identified some sufficient or necessary initial conditions for obtaining the wild
type steady state (minimal pre-patterns). 
We now analyze how mutant patterns arise from gene ``knockout'' experiments or delay in establishing the 
initial pre-pattern.

Gene knockout experiments consist of completely supressing the expression of a given gene in all cells.
In our models this is equivalent to setting the corresponding mRNA permanently zero in all equations.
Thus a {\it wingless} knockout can be analyzed by setting $\wg_i=0$  for all $i=1,2,3,4$, and
in every equation of the Boolean rules in Table~\ref{table_rules}.
The steady states of the resulting system can now be computed from: 
\beqn{eq-knockoutF}
  X^* = F^*(X^*),
\eeqn
where the vector $X^*$ and functions $F^*$ include all but the knockout variables.
For example, for the {\it wingless} knockout, we drop the functions $F_{\twg_i}$ and for
the other mRNAs and proteins we have
\beq
  &&  \WG_i=F^*_{\tWG_i} = 0, \ \  i=1,2,3,4   \\
  &&  \mbox{node }=F^*_{\mbox{\tiny node}} = F_{\mbox{\tiny node}}, \ 
      \mbox{ for nodes } \en,\hh,\ci,\ptc,\EN,\HH,\PTC,\CI,\CIA,\CIR,
\eeq 
because $\wg$ appears explicitly only in the rules $F_{\tWG_i}$.
So, it follows immediately that $\WG_i=0$, and the remaining nodes are then 
also easy to compute.
It is easy to check (see also~\cite{ao03}) that knockouts of $\wg$, $\en$, $\hh$ or $\ptc$
exhibit only one steady state, while knockouts of $\ci$ exhibit three steady states (summarized
in Table~\ref{tab-knockouts}) in both asynchronous and Glass-type models.

\begin{table}
\caption{The steady states corresponding to gene knockouts 
in the segment polarity network model, calculated according to\rf{eq-knockoutF}.
Here, for the $\ci$ knockout, the ``wild type'' state is interpreted 
as the wild type pattern in all but the $\ci$, $\CI$, $\CIA$,$\CIR$  mRNA/proteins.}
\label{tab-knockouts}
\begin{center}
\begin{tabular}{cc}
\hline
  Knockouts              &  Mutant steady states \\
\hline\hline
  $\wg$, $\en$, $hh$    &   no segmentation \\
\hline
  $\ptc$                &   broad stripes \\
\hline
                &   wild type, \\
   $\ci$        &  broad stripes , \\
                &   ectopic \\
\hline
\end{tabular}
\end{center}
\end{table} 

From another point of view, one may consider a delay in the establishment of the pre-pattern
(that is, the full initial condition\rf{eq-initial-wt}). If expression of a given gene is delayed, 
does the system recover, and how soon?
To answer this question, we simulated a delay in expression of gene $X$, by
setting the corresponding discrete variable $X(t)=0$ in all cells, 
for all $t\leq T_{\mbox{\tiny delay}}$. 
We then varied $T_{\mbox{\tiny delay}}$ between 0 and 7 time units, and measured the frequency of 
occurrence of each steady state, both for the asynchronous and Glass-type models.
(In the latter we set the concentration threshold $\theta$ equal to $1/2$ for all nodes.)

\begin{figure}
\centerline{
\psfig{figure=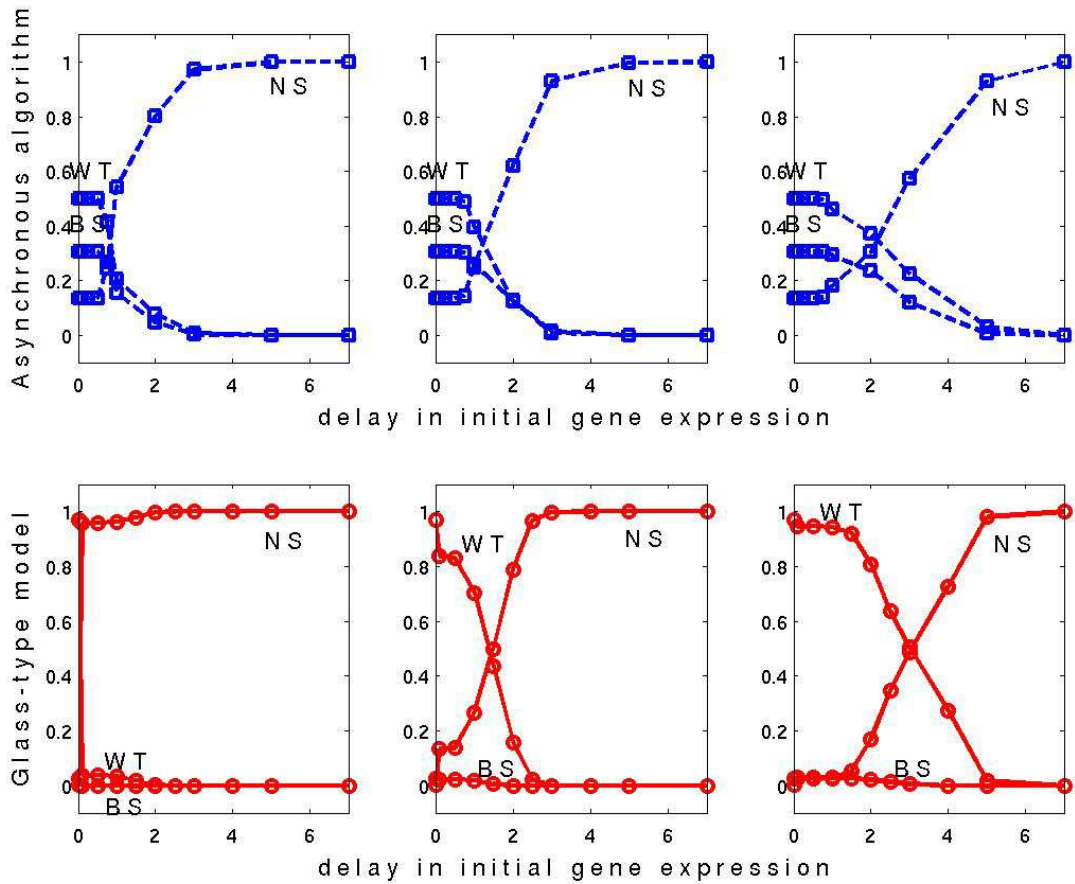,height=5in} }
\caption{Effect of initially delayed expression in occurrence of steady state patterns.
Left: delay in {\it wingless} expression. Middle: delay in {\it engrailed} expression. 
Right: delay in {\it hedgehog} expression. 
(results out of 1000 runs).}
\label{fig-delay-wg-en-hh}
\end{figure}

\begin{figure}
\centerline{
\psfig{figure=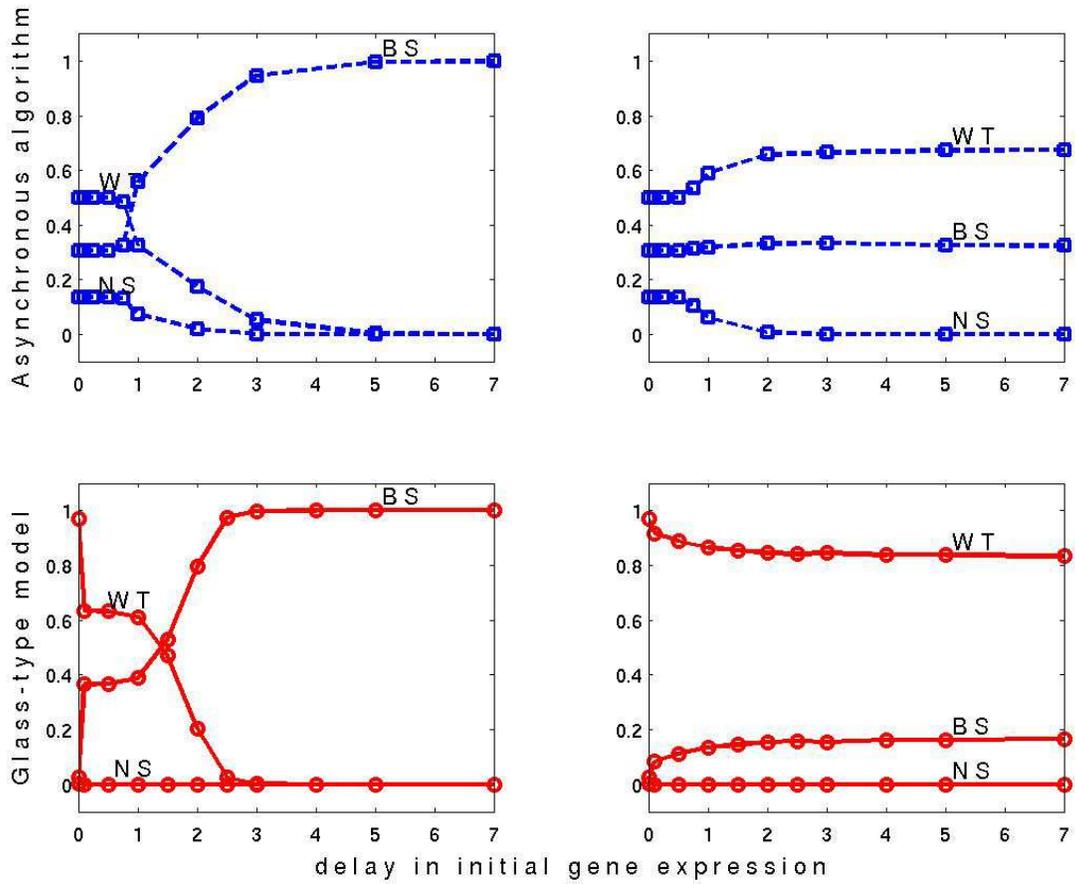,height=5in} }
\caption{Effect of initially delayed expression in occurrence of steady state patterns.
Left: delay in {\it patched} expression. Right: delay in {\it cubitus interruptus} expression. 
(results out of 1000 runs).}
\label{fig-delay-ptc-ci}
\end{figure}

The results are shown in Figures~\ref{fig-delay-wg-en-hh} and~\ref{fig-delay-ptc-ci}.
We can see that, for short $T_{\mbox{\tiny delay}}$, both models recover their original
frequencies of occurrence of each steady state, while for long $T_{\mbox{\tiny delay}}$ 
both models converge to the corresponding mutant steady state. 
A curious exception is the case of $\ci$, where long delays in its initial expression do not significantly
change the probability that the wild type steady state is achieved (and even slightly increase it
in the asynchronous model). This agrees with the conclusion of~\cite{ao03} that {\it cubitus interruptus} knockout
coupled with an otherwise wild-type initial condition converges to a state close
to the wild type steady state. Remarkably, delays in both cubitus interruptus transcription factors
(CIA, CIR) have a lesser effect than an imbalance in their expression (see~\cite{cas05}). This leads us to 
predict that, during the pre-segmentation stage of embryo development, the cubitus interruptus proteins'
expression is the last to be established.

Another noteworthy observation is the fact that small delays in $\wg$ expression have much more drastic 
effects on the system than the same delay in $\en$ or $\hh$.
This phenomenon reflects the one-way signaling cascade starting with expression of $\wg$, 
which induces $\en$, which in turn induces $\hh$.
We see that a total disruption in the system is caused by a delay of only 3 time units in $\wg$ 
or $\en$ expression, which cause the system to fail to reproduce the wild type pattern, and
settle into a non-segmented pattern. However, if only $\hh$ is delayed, the system
is disrupted only after a delay of 5 time units.
In other words, recovery of the system back to the ``good'' developmental process is more probable in 
the event of a $\hh$ expression delay, than a $\wg$ or $\en$ expression delay.

\section{Robustness of the model under timescale separation}
\label{sec-timesep}
 
In the previous algorithms, the space of all possible timescales for protein/mRNA regulatory
processes was explored, with no assumptions on the characteristic duration of translational
or post-translational processes.
As a consequence, the robustness analysis shows that the model diverges from the wild type 
pattern very often, with the biologically inviable states occurring with a noticeable 
frequency.
However, it is also well known that post-translational processes such as protein 
conformational changes or complex formation, usually have shorter durations than transcription, 
translation or mRNA decay.
This fact justifies the introduction of a distinct timescale separation among processes, 
by choosing to update proteins first and mRNAs later.

\subsection{Timescale separation in the random order algorithm}
Timescale separation is straightforwardly implemented in the random order algorithm
presented in Section~\ref{sec-asynch}; at the $k$-th updating step we generate
two random permutations, $\phi^k_{\tProt}$ and $\phi^k_{\tmRNA}$,
within the set of proteins and mRNAs, respectively. Then the $N$ nodes are updated in the order 
given by 
\beq
  \phi^k=(\phi^k_{\tProt},\;\phi^k_{\tmRNA}).
\eeq
{This method again shows that the Boolean model is very robust, in the sense that 
when started from the wild type initial condition, the wild type 
pattern occurs with a frequency of $87.5\%$ and only one other steady state is observed, the 
broad striped pattern, with a frequency of $12.5\%$.}
Furthermore, these frequencies are exact as we show in~\cite{cas05}, where we also completely 
characterize the model resulting from incorporation of a protein/mRNA timescale separation 
into the random order algorithm.
We show that the wild type state is in fact an attractor for the system, 
while the pathway to the broad stripes state may exhibit oscillatory cycles.
We summarize this and other results in the next theorem, stated without proof, and refer 
to~\cite{cas05} for more details.

\begin{theo}{th-random}
In the random order algorithm with timescale separation, 
let $\wg_3^0=0$, $\ptc_3^0=1$, $\hh_{2,4}^0=0$ and $\ci_3^0=1$ 
(as satisfied by the initial pattern\rf{eq-initial-wt}).
Then system diverges from the wild type pattern if and only if the 
permutation $\phi^1$ satisfies the following sequence among the proteins 
$\CI$, $\CIA$, $\CIR$ and $\PTC$:  
\beqn{eq-perm-wg3=1}
\begin{array}{cccccc}
    \CIR_3 & \CI_3 & & \CIA_3  & & \PTC_3,        \\
%    & & & & & \\ 
          & \CI_3 &   \CIR_3  & \CIA_3 & & \PTC_3,  \\
%    & & & & & \\ 
          & \CI_3 & & \CIA_3  & \CIR_3 & \PTC_3,
\end{array}
\eeqn
while the other proteins may appear in any of the remaining slots.
\bbox
\end{theo}

Thus we can compute the exact probability with which the random order algorithm (with 
timescale separation) leads to either the wild type or broad stripes pattern:
the latter is simply the fraction of sequences of the form\rf{eq-perm-wg3=1},
and equals $12.5\%$~\cite{cas05}.

\subsection{Timescale separation in the Glass-type and asynchronous algorithms}
As shown in Section~\ref{sec-glass}, the (discrete) asynchronous algorithm and the (piecewiese continuous)
Glass-type system provide equivalent representations of a gene expression network. Indeed, 
the ``specific time units'' $\gamma_i$ and the inverse ``scaling factors'' $\alpha_i^{-1}$,
both represent the rate of dynamical evolution of each individual node.
For these two models, we implement time separation among processes by using two 
non-overlapping intervals for the scaling factors:
\beq
&&  \gamma_i^{-1},\alpha_i \in A_{\tmRNA},\ \ \ \mbox{if }\ X_i\in\{\wg,\en,\hh,\ptc,\ci \} \\
&&  \gamma_i^{-1},\alpha_i \in A_{\tProt},\ \ \ \mbox{if }\ X_i\in\{\WG,\EN,\HH,\PTC,\CI,\CIA,\CIR\},
\eeq
with, for instance, $A_{\tmRNA}=[0.2,0.6]$ and $A_{\tProt}=[1.4,1.8]$.
Under these conditions, choosing the factors $\alpha_i$ from a uniform distribution in these
intervals, numerical experiments indicate that the two methods respond in mostly similar ways,
with only two patterns occurring at steady state when the systems
start from (wild type) initial condition\rf{eq-initial-wt}. 
The two possible steady states are the wild type and broad stripes patterns.

For the asynchronous algorithm, the probabilities of convergence to
each of the steady states clearly depend on the {\it distance} between the two intervals:
convergence to wild type is $93\%$ in the case where intervals $A_{\tmRNA}$, $A_{\tProt}$
are consecutive, and up to $100\%$ for the case $2a<b$, $a\in A_{\tmRNA}$, $b\in A_{\tProt}$.

\begin{figure}
\centerline{
\psfig{figure=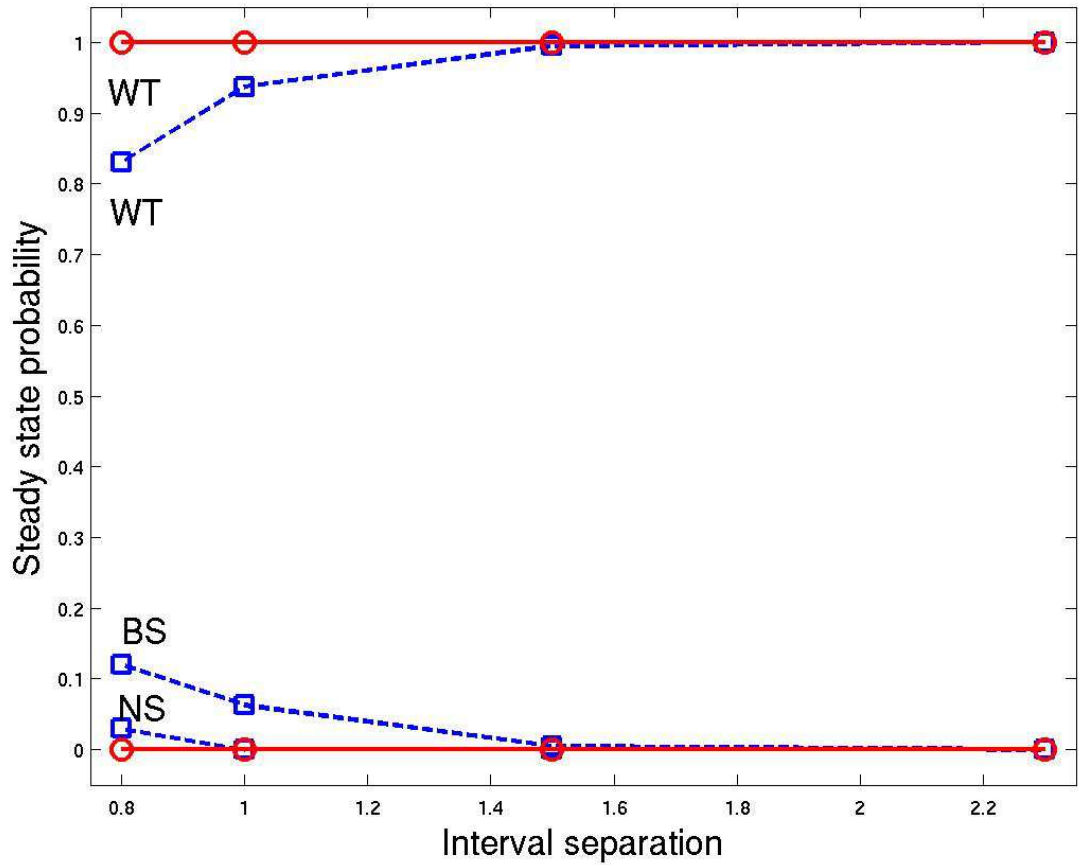,height=5in} }
\caption{Probability of occurrence of the three most frequent patterns:
wild type (WT), broad stripes (BS), and no segmentation (NS), with separation of timescales. 
Dashed lines/squares represent asynchronous algorithm results, while solid lines/circles 
represent Glass-type model results. The $x$-axis represents the level of separation, 
computed by $\min\{b\in A_{\tProt}\}/\max\{a\in A_{\tmRNA}\}$  (out of 1000 runs).
Results were obtained with $\theta=0.5$.}
\label{fig-time-separation}
\end{figure}

For the Glass-type model, two cases can be distinguished: $\theta\leq0.5$ and $\theta>0.5$. 
For $\theta\leq0.5$, numerical simulations show that the
model reaches wild type pattern with probability near 100\%, even when there
is some overlap between $A_{\tmRNA}$ and $A_{\tProt}$ (see Fig.~\ref{fig-time-separation}).
In fact, we next theoretically prove that {\em the wild type pattern is indeed 
the unique possible steady state} of the hybrid system\rf{eq-glass-a}
and initial condition\rf{eq-initial-wt}, as indicated by
the simulations, when there is a suitable distance between the intervals,
and a lower bound on $\theta$ (Theorem~\ref{th-glass1}, Section~\ref{sec-convergence}). 
For $\theta>0.5$, we have found no condition that guarantees convergence to the
wild type steady state, and indeed numerical simulations show that, even for large interval
separation, the system may converge to one of the mutant patterns.

A comparison of Theorems~\ref{th-random} and~\ref{th-glass1} 
emphasizes differences and similarities between discrete and continuous models: 
intuitively, the single discrete event described by Theorem~\ref{th-random} cannot take 
place in a continuous model. Therefore $\wg_3$ remains ``0'' (OFF) for all times, ruling out 
the possibility that the broad stripes pattern is reached.
Indeed, Theorem~\ref{th-random} establishes that (in the discrete
case, with the random order algorithm) 
divergence from wild type pattern occurs if and only if $\wg_3^1=1$. 
This fact involves a jump in $\wg_3$ from ``0'' to ``1'' at precisely 
the first iteration. 
On the other hand, in the Glass-type model,  the continuous variable $\cwg_3$ cannot 
instantaneously jump from ``0'' to ``1''.
Since the discrete ON/OFF levels are defined by a threshold on $\cwg_3$, there will
necessarily be a smoothing effect on any transition between ``0'' and ``1''.
This is what happens in case (a) of Theorem~\ref{th-glass1}.

The second (sufficient) condition of Theorem~\ref{th-glass1} guarantees convergence to
the wild type steady state for all $0<\theta\leq0.5$  (while condition (a) is only for 
$0.382\leq\theta\leq0.5$), but assumes that $\alpha_{\tPTC_3}>\alpha_{\tCI_3}$.
This is an analog to Theorem~\ref{th-random}: if $\alpha_{\tPTC_3}>\alpha_{\tCI_3}$,
then (starting from $\PTC_3(0)=\CI_3(0)=0$ and assuming $F_{\tPTC_3}=F_{\tCI_3}=1$)
$\cPTC_3$ increases faster than $\cCI_3$, implying that $\PTC_3$ becomes ON faster
than $\cCI_3$. Such response prevents the events listed in Theorem~\ref{th-random},
which would lead to a mutant state.
Thus, both discrete and piecewise linear model predict that the sequence of $\PTC$,
$\CI$ expression in the third cell is one of the fundamental pieces in establishing
the correct development of embryo segmentation.

It is also worth pointing out that the three methods provide qualitatively similar
results under timescales separation, all predicting that only wild type and broad stripes 
patterns to occur, the latter with considerably smaller frequency.
\begin{table}
\caption{Probabilities of convergence to a given steady state,
under the separation of timescales assumption.
These values are theoretically exact for both the random order algorithm and Glass-type model.}
\label{tab-two-timescales}
\begin{center}
\begin{tabular}{lccc}
\hline
 Steady state      & Random order & Asynchronous & Glass-type\\
   pattern         & algorithm & algorithm &      model\\
\hline\hline
  wild type      &  87.5\% & 93.7-100\% &  100\% \\
%\hline
  broad stripes   &  12.5\% & 0-6.3\% &  0\% \\
%\hline
  other          &  0\% & 0\%  & 0\% \\
\hline 
\end{tabular}
\end{center}
\end{table} 

\subsection{Distinct concentration thresholds for ON state}
\label{sec-theta-vary}
A natural question arising in the analysis of equation\rf{eq-glass-a} concerns
the dynamics of the system under more general concentration thresholds.
We have seen that $\theta$ (even when equal for all species) plays an important role 
in establishing basins of attraction to each of the steady states of the model.

We have, in particular, identified three distinct regions of behavior:
\beq
  \begin{array}{lc}
    \mbox{Region 1}: &  0<\theta<\frac{3-\sqrt{5}}{2}\\
    & \\
    \mbox{Region 2}: & \frac{3-\sqrt{5}}{2}\leq\theta\leq\frac{1}{2}\\
    & \\
    \mbox{Region 3}: & \frac{1}{2}<\theta\leq1
  \end{array}
\eeq 
When $\theta$ is equal for all nodes, the probability that the system evolves into 
the wild type pattern is above 90\% in regions 1 and 2, but may be as low as 68\% 
in region 3.
Furthermore, for region 2, we theoretically prove that that probability is exactly 100\%
under the timescale separation assumption.

We now associate to each node a specific $\theta_i$, so that\rf{eq-threshold} is modified to:
\beqn{eq-threshold2}
   X_i(t)=\left\{ \begin{array}{ll}
                          0, & \cX_i(t)\leq\theta_i \\
                          1, & \cX_i(t)>\theta_i\ .
               \end{array}
       \right.                   
\eeqn
To test the performance of the system and compare it to previous results, 
we considered two timescale situations: $\alpha_i\in[0.5,1.5]$ for all $i$,
or the timescale separation $A_{\tmRNA}=[0.2,0.6]$, $A_{\tProt}=[1.4,1.8]$.
In each case, we randomly assigned values to $\theta_i$ from 
uniform distributions in the intervals $(0,1)$, $(0,0.5)$ and $(0.4,0.5)$.
(Note that $0.4$ is close to $(3-\sqrt{5})/2$.)
The numerical results with varying $\theta_i$ extend and confirm our previous observations
for $\theta_i=\theta$, $i=1,\ldots,N$.

Table~\ref{tab-theta-vary} summarizes the results for all combinations of $\theta_i$ and 
$\alpha_i$ regions.
The most general case, allowing a large degree of freedom in both timescales and concentration
thresholds, indicates the vulnerability of the network, with a very high incidence on mutant 
patterns ($\theta_i\in(0,1)$, $\alpha_i\in[0.5,1.5]$).
Again we see that there is a marked difference in the $\theta_i$ regions below or above $0.5$.
Comparing the reasonable results obtained for $\theta_i\leq0.5$ with the bad performance for
$0.5<\theta_i<1$, we conclude that the optimal ON concentratin for proteins or mRNA is below 50\%
of maximal concentration.

Restricting $\theta_i$ even further to one of the conditions given in Theorem~\ref{th-glass1}
(Section~\ref{sec-convergence}; conditions developed for the case $\theta_i=\theta$, $i=1,\ldots,N$)
dramatically increases the probability that the system develops in the correct way.

These simulations suggest, moreover,  that some of the theoretical results obtained for each of the 
three regions may extend to the case of distinct $\theta_i$.
Indeed, note that, under timescale separation,
when $\theta_i$ are chosen from region 2, convergence to wild type steady state is 100\%
-- compare to part (a) of Theorem~\ref{th-glass1}.

These simulations further confirm the role of $\alpha_{\tPTC_3}$ and $\alpha_{\tCI_3}$
in the segmentation network. 
Requiring $\alpha_{\tPTC_3}>\alpha_{\tCI_3}$ decreases the probability of formation of the broad stripes
pattern (any timescales), but doesn't influence the probability of a non segmented embryo. The latter
mutant is prevented only by a complete separation of timescales of the regulatory processes.

\begin{table}
\caption{Probabilities of convergence to a given steady state,
with distinct concentration thresholds $\theta_i$ and distinct timescales $\alpha_i$,
in the Glass-type model. (Probabilities computed out of 1000 simulations for each case.)}
\label{tab-theta-vary}
\begin{center}
\begin{tabular}{lcccccc|r}
\hline
 Steady state      &  $(0,1)$  & $(0.5,1)$ & $(0,0.5]$  & $(0,0.5]$ & $[0.4,0.5]$ & $0.5$  & \multicolumn{1}{l}{$\theta_i$} \\                          
    pattern        &   &          &            & $\alpha_{\tPTC_3}>\alpha_{\tCI_3}$ & & &  $\alpha_i$  \\
\hline\hline
  wild type           &  45.6\% & 57.1\% & 84.1\% &  90\% & 92.6\% & 94.2\%      & $A_{\tmRNA}=$ \\
%\hline
  broad stripes       &  27.8\% & 15.1\% & 12\% &  6.2\% & 7.3\%    & 4.5\%      & $A_{\tProt}=$ \\
%\hline
  no segmentation     &  24.4\% & 25.8\% & 0.9\%  & 0.9\% & 0.05\%   & 1.3\%      &   $[0.5,1.5]$\\
%\hline
  wild type, variant  &  2.1\% & 1.9\% & 2.9\%  & 2.9\% & 0\%     & 0\%      &   \\
\hline\hline
  wild type           &  74.1\% & 52.7\% & 96.6\% &  97.1\% & 100\% & 100\%    & $A_{\tmRNA}=$    \\
%\hline
  broad stripes       &  10.8\% & 3.3\% & 3.3\% &  2.8\% & 0\%      & 0\%    & $[0.2,0.6]$\\
%\hline
  no segmentation     &  14.1\% & 43.9\% & 0\%  & 0\% & 0\%         & 0\%    & $A_{\tProt}=$ \\
%\hline
  wild type, variant  &  1.0\% & 0\% & 0.1\%  & 0.1\% & 0\%         & 0\%    & $[1.4,1.8]$\\
\hline
\end{tabular}
\end{center}
\end{table}

\subsection{Glass-type model provides exact convergence to wild type pattern}
\label{sec-convergence}
In this section we will require that the intervals $A_{\tmRNA}$ and $A_{\tProt}$ 
do not overlap, by satisfying the following assumption:
\beqn{eq-separation}
 \mbox{For all }\ \  a\in A_{\tmRNA}\ \ \mbox{and}\ \  b\in A_{\tProt}:\ \ \ 0<2a<b.
\eeqn
A second assumption is that the effective maximal concentration is equal for all nodes and 
satisfies $\theta\leq1/2$, which is equivalent to:
\beqn{eq-lowthres}
  \ltho\leq\lth
\eeqn

\begin{theo}{th-glass1}
Consider system\rf{eq-glass-a} with initial condition\rf{eq-initial-wt}. 
Assume that the scaling factors $\alpha_i$ satisfy\rf{eq-separation}.
Assume also that one of the following conditions holds:
\bit 
\item[(a)]
$\theta\leq1/2$ and $(1-\theta)^2\leq\theta$ 
or equivalently $0.382\approx(3-\sqrt{5})/2\leq\theta\leq1/2$;

\item[(b)] 
$\theta\leq1/2$ and $\alpha_{\tPTC_3}>\alpha_{\tCI_3}$;
\eit
then $\wg_3(t)=0$ for all $t$.
\end{theo}
This shows that the steady state representing the broad stripes pattern
cannot ever be reached in system\rf{eq-glass-a} from the initial
condition\rf{eq-initial-wt}, when $\theta\leq1/2$ and either of the extra 
conditions holds.

\begin{theo}{th-glass2}
Consider system\rf{eq-glass-a} with initial condition\rf{eq-initial-wt}. 
Assume that the scaling factors $\alpha_i$ satisfy\rf{eq-separation}, and
that $\theta\leq1/2$. 
Then  $\wg_4(t)=1$ and $\PTC_1(t)=0$ for all $t$.
\end{theo}
This shows that the steady states represented by the no segmentation, wild type variant or the two ectopic patterns
also cannot ever be reached in system\rf{eq-glass-a} from the initial condition\rf{eq-initial-wt}.
From Theorems~\ref{th-glass1} and~\ref{th-glass2}
we conclude that, under the timescale separation assumption, the
Glass-type model\rf{eq-glass-a} can only converge to the wild type pattern,
when starting from the initial condition\rf{eq-initial-wt}, and for appropriate $\theta$ values
(Table~\ref{tab-two-timescales}).

We first summarize some useful observations. 
Let $X$ denote any of the nodes in the network, and $\alpha$ its time rate.
Since equations\rf{eq-glass-a} are either of the form $d\cX/dt=\alpha(-\cX+1)$
or $d\cX/dt=-\alpha\cX$, their solutions are continous functions,
piecewise combinations of:
\beqn{eq-hybrid-sol1}
  \cX^1(t) &=& 1 - (1-\cX^1(t_0))\,e^{-\alpha(t-t_0)} \\
  \cX^0(t) &=& \cX^0(t_0)\,e^{-\alpha(t-t_0)} \label{eq-hybrid-sol0}
\eeqn
$\cX^1(t)$ (resp. $\cX^0(t)$) is monotonically increasing (resp. decreasing).
In addition, note that discrete variables $X$ can only switch between 0 and 1 
at those instants when $\cX(t_{\mbox{\tiny switch}})=\theta$, that is:
\beqn{eq-hybrid-switch1}
  t_{\mbox{\tiny switch}}^{1} &=& t_0 +\frac{1}{\alpha} \,\lthow{(1-\cX(t_0))}\\
  t_{\mbox{\tiny switch}}^{0} &=& t_0 +\frac{1}{\alpha} \,\lthw{\cX(t_0)}  
        \label{eq-hybrid-switch0}
\eeqn

From the initial conditions, together with the constant values of $\SLP_i$ ($i=1,2,3,4$), we 
can immediately conclude:
\beqn{eq-wgen}
 && \cwg_{1,2}(t)=\WG_{1,2}(t)=0,\\
 && \cen_{3,4}(t)=\cEN_{3,4}(t)=0,  \nonumber \\
 && \chh_{3,4}(t)=\cHH_{3,4}(t)=0, \label{eq-wgen2}
\eeqn
for all $t\geq0$. 
Then, because $\ci_{3,4}(0)=1$ and $F_{\tci_{3,4}}=\mbox{not }\EN_{3,4}$,
\beqn{eq-ciCI}
 \cci_{3,4}(t)=1\ \mbox{ and }\ \cCI_{3,4}(t)=1-e^{-\alpha_{\tCI_{3,4}}\, t}.
\eeqn

\begin{lem}{lm-wg3} 
Let $0\leq t_0<t_3\leq t_1$ and $0\leq t_2<t_3$. 
Define $\delta=\ltho/\max_{1,\ldots,N}{\alpha_i}$.
Assume $\CIA_3(t)=0$ for $t\in(t_2,t_3)$, and $\wg_3(t)=0$ for $t\in[0,t_3)$.
Then
\bit
\item[(a)] $\wg_3(t)=0$ for $t\in[0,t_3+\delta)$;
\item[(b)] $\WG_3(t)=0$ for $t\in[0,t_3+\delta)$;
\item[(c)] $\en_{2}(t)=\EN_{2}(t)=0$ for $t\in[0,t_3+\delta)$;
\item[(d)] $\hh_{2}(t)=\HH_{2}(t)=0$ for $t\in[0,t_3+\delta)$.
\eit
Assume further that $\PTC_3(t)=1$ for $t\in(t_0,t_1)$. Then
\bit
\item[(e)] $\PTC_3(t)=1$ for all $t\in(t_0,t_3+\delta)$.
\item[(f)] $\CIA_3(t)=0$ for all $t\in(t_2,t_3+\delta)$.
\eit
\end{lem}

\bprf
Part (a) follows directly from the fact that $F_{\twg_3}(t)=0$ on $[0,t_3)$, and
from\rf{eq-hybrid-switch1}.

To prove parts (b), (c), and (d), first note that initial conditions together with 
$\wg_3(t)=0$ for $t\in[0,t_3)$ imply
\beq
 && \cWG_3(t)=0,\ \cen_{2}(t)=\cEN_{2}(t)=0,\\
 && \chh_{2}(t)=\cHH_{2}(t)=0,
\eeq
for $t\in[0,t_3]$. Then, from equations\rf{eq-hybrid-sol1} to\rf{eq-hybrid-switch0}
we conclude that the corresponding discrete variables cannot switch from 0 to 1 during  
an interval of the form $[0,t_3+\frac{1}{\alpha_j}\ltho)$. Taking the largest common interval
yields the desired results.

To prove parts (e) and (f), assume also that $\PTC_3(t)=1$ for $t\in(t_0,t_1)$. 
From\rf{eq-wgen2} and part (d), it follows that function $F_{\tPTC_3}$ does not switch 
in the interval $(t_0,t_3+\delta)$ and in fact $\PTC_3(t)=1$ for all $t$ in this interval.
This, together with\rf{eq-wgen2} and part (d) yield $F_{\tCIA_3}(t)=0$ for $(t_0,t_3+\delta)$, 
so that $\cCIA_3$ cannot increase in this interval and the discrete level satisfies
$\CIA_3(t)=0$ for all $t\in(t_2,t_3+\delta)$, as we wanted to show.
\bbox

\begin{cor}{cr-wg3} 
Let $0\leq t_0<t_3\leq t_1$ and $0\leq t_2<t_3$. 
If $\PTC_3(t)=1$ for $t\in(t_0,t_1)$, 
$\CIA_3(t)=0$ for $t\in(t_2,t_3)$, and $\wg_3(t)=0$ for $t\in[0,t_3)$,
then $\wg_3(t)=0$ for all $t$.
\end{cor}

\bprf
Applying Lemma~\ref{lm-wg3} we conclude that, given any $k\geq0$:
\beq
  && \CIA_3(t)=0,\ \mbox{ for }\ t\in(t_2,t_3+k\delta)\\
  &&  \wg_3(t)=0,\  \mbox{ for }\ t\in[0,t_3+k\delta)\\
  &&  \PTC_3(t)=1\ \mbox{ for }\ t\in(t_0,t_3+k\delta) 
\eeq
imply 
\beq
  &&  \CIA_3(t)=0,\ \mbox{ for }\ t\in(t_2,t_3+(k+1)\delta)\\
  &&  \wg_3(t)=0,\  \mbox{ for }\ t\in[0,t_3+(k+1)\delta)\\
  &&  \PTC_3(t)=1\ \mbox{ for }\ t\in(t_0,t_3+(k+1)\delta). 
\eeq
Since $\delta$ is finite, we conclude by induction on $k$
that $\wg_3(t)=0$ for all $t$.
\bbox

{\it Proof of Theorem~\ref{th-glass1}:}
The rule for $\CIA_3$ may be simplified to (by\rf{eq-wgen2}) 
\beq
  F_{\tCIA_3}=\CI_3\mbox{ and }[\mbox{not}\PTC_3\mbox{ or }\hh_2 \mbox{ or }\HH_2].
\eeq
From equation\rf{eq-ciCI}, we have that 
\beqn{eq-CI3}
   \CI_3(t)=1,\ \mbox{ for all } t>\frac{1}{\alpha_{\tCI_3}}\ltho. 
\eeqn
On the other hand, since $\ptc_3(0)=1$, by continuity of solutions
$\ptc_3(t)=1$ for all $t<\frac{1}{\alpha_{\tptc_3}}\lth$.
This implies that the Patched protein satisfies
\beq
   \cPTC_3(t)=1-e^{-\alpha_{\tPTC_{3}}\, t},\ \ 0\leq t\leq\frac{1}{\alpha_{\tptc_3}}\lth
\eeq
and therefore 
\beqn{eq-PTC3}
   \PTC_3(t)=\left\{
            \begin{array}{ll}
               0, & 0\leq t\leq\frac{1}{\alpha_{\tPTC_3}}\ltho\\
               1, & \frac{1}{\alpha_{\tPTC_3}}\ltho<t<\frac{1}{\alpha_{\tptc_3}}\lth.
            \end{array}
            \right.     
\eeqn
By assumption, $\alpha_{\tPTC_3}>\alpha_{\tptc_3}$ and also $\ltho\leq\lth$, 
defining a nonempty interval where $\PTC_3$ is expressed. 
Now let $t_c=\frac{1}{\alpha_{\tCI_3}}\ltho$ and $t_p=\frac{1}{\alpha_{\tPTC_3}}\ltho$.
$\cCIA_3(t)$ starts at zero and must remain so while  $\CI_3=0$, so that 
\beq
  \CIA_3(t)=0\ \mbox{ for } 0<t<t_c.
\eeq
In the case $t_c>t_p$, 
letting $t_0=t_p$, $t_1=\frac{1}{\alpha_{\tptc_3}}\lth$, $t_2=0$, and $t_3=t_c$
in Corollary~\ref{cr-wg3}, obtains $\wg_3(t)=0$ for all $t$.
This proves item (b) of the theorem, and part of (a). 

To finish the proof of item (a), we assume that $(1-\theta)^2<\theta$ and must now consider the 
case $t_c\leq t_p$.
Then
\beq
  \cCIA_3(t)=\left\{
             \begin{array}{ll}
                0, & 0\leq t\leq t_c   \\
                1-e^{-\alpha_{\tCIA_3}\,(t-t_c)}, &   t_c < t \leq t_p \\
                \cCIA_3(t_p)\,e^{-\alpha_{\tCIA_3}\,(t-t_p)}, &
                                      t_p < t\leq \frac{1}{\alpha_{\tptc_3}}\lth,
             \end{array}\right.
\eeq
Following equation\rf{eq-hybrid-switch1} with $t_0=t_c$ and $\cCIA_3(t_0)=0$,
$\CIA_3$ might become expressed at time $t_c<t_a<t_p$:
\beq
   t_a=t_c+\frac{1}{\alpha_{\tCIA_3}}\ltho,
\eeq
but it would then become zero again at (equation\rf{eq-hybrid-switch0} with
$t_0=t_p$)
\beq
   t_b=t_p + \frac{1}{\alpha_{\tCIA_3}}\lthw{\cCIA_3(t_p)}.
\eeq
Finally, we show that, even if $\CIA_3(t)=1$ for $t\in(t_a,t_b)$,  $\wg_3$ cannot become  
expressed in this interval. In this interval, $\cwg_3$ evolves according to
$
  \cwg_3(t)=1-e^{-\alpha_{\twg_3}\,(t-t_a)},
$
and $\wg_3$ can switch to 1 at time 
\beq
  t_w=t_a+\frac{1}{\alpha_{\twg_3}}\ltho.
\eeq
We will show that $t_w>t_b$, so $\wg_3(t)=0$ in the interval $[0,t_b)$. 
Writing 
\beq
  \lthw{\cCIA_3(t_p)}&=&\lthow{\cCIA_3(t_p)}\frac{1-\theta}{\theta} \\
   &=&\lthow{\cCIA_3(t_p)} +\ln\, \frac{1-\theta}{\theta} \\
   &\leq&\ltho +\ltho
\eeq
where we have used $\cCIA_3(t_p)\leq1$ and the assumption on 
$\theta$: $\frac{1-\theta}{\theta}\leq\frac{1}{1-\theta}$.
Therefore
\beq
  t_b & \leq & t_p +\frac{2}{\alpha_{\tCIA_3}}\ltho \\
      & < & \frac{1}{\alpha_{\twg_3}}\ltho+\frac{1}{\alpha_{\tCIA_3}}\ltho<t_w
\eeq
where we have used the timescale separation assumption\rf{eq-separation}.
Letting $t_0=t_p$, $t_2=0$, and $t_1=t_3=\min\{t_b,\alpha_{\tptc_3}^{-1}\lth\}$ in the Corollary,
obtains $\wg_3(t)=0$ for all $t$.
\bbox

We will next show that if $\wg_4(t)=1$ in a given interval $[0,T)$, 
then in fact $\wg_4(t)$ remains expressed for a longer time, up to $T+\delta$,
with $\delta>0$.
This is mainly due to assumption\rf{eq-separation}, which says that mRNAs take longer 
than proteins to update their discrete values, because they have longer half-lives: 
$\alpha_{\tmRNA}^{-1}>\alpha_{\tProt}^{-1}$. 
This allows the initial signal ``$\wg_{4}=1$'' to travel down the network, sequencially
affecting the wingless protein, {\it engrailed, hedgehog} and CIA, and feed back into
{\it wingless} 
allowing $\wg_4$ to remain expressed for a further time interval.

\begin{lem}{lm-wg4}
Let $T\geq \frac{1}{\alpha_{\twg_4}}\lth$ and define 
\beqn{eq-delta}
  \delta=\frac{1}{\alpha_{\tWG_4}}\,\lthw{ (1-e^{-\frac{\alpha_{\tWG_4}}{\alpha_{\twg_4}}\lth})}.
\eeqn
If $\wg_4(t)=1$ for $0\leq t< T$, then
\bit
\item[(a)] $\WG_4(t)=1$ for $t\in(\frac{1}{\alpha_{\tWG_4}}\ltho,T+\delta)$;
\item[(b)] $\en_1(t)=1$ for $t\in[0,T+\delta)$;
\item[(c)] $\cEN_1(t)=1-e^{-\alpha_{\tEN_1}t}$ for $t\in[0,T+\delta)$, 
           and $\EN_1(t)=1$ for $(\frac{1}{\alpha_{\tEN_1}}\ltho,T+\delta)$;
\item[(d)] $\ci_1(t)=0$, $\CI_1(t)=0$, $\CIA_1(t)=0$, and $\CIR_1(t)=0$ for $t\in[0,T+\delta)$;
\item[(e)] $\hh_1(t)=1$, for $t\in[0,T+\delta)$;
\item[(f)] $\CIA_4(t)=1$, for $t\in(\frac{1}{\alpha_{\tCI_4}}\ltho
                                    +\frac{1}{\alpha_{\tCIA_4}}\ltho,T+\delta)$,
           and $\CIR_4(t)=0$, for $t\in[0,T+\delta)$;
\item[(g)] $\wg_4(t)=1$ for $t\in[0,T+\delta)$.
\eit
\end{lem}

\bprf 
Let $T\geq\frac{1}{\alpha_{\twg_4}}\lth$, and assume that $\wg_4(t)=1$ for $0\leq t< T$. 
To prove part (a), note that $\cWG_4(t)$ is of the form\rf{eq-hybrid-sol1} 
(with $t_0=0$, and $\cWG_4(0)=0$) and the corresponding discrete variable is 
$\WG_4(t)=1$, for $t\in(\frac{1}{\alpha_{\tWG_4}}\ltho,T)$.
Moreover, suppose that $\wg_4(t)=0$ for $t>T$, then
\beq
  \cWG_4(t)=(1-e^{-\alpha_{\tWG_4}T})e^{-\alpha_{\tWG_4}(t-T)},\ \ \ t>T.
\eeq
But $\WG_4$ remains 1 until the switching threshold is attained, that is up to time
\beq
 & &   T+\frac{1}{\alpha_{\tWG_4}}\lthw{ (1-e^{-\alpha_{\tWG_4}T})}\\
 &\geq & 
   T+\frac{1}{\alpha_{\tWG_4}}\lthw{ (1-e^{-\alpha_{\tWG_4}\frac{1}{\alpha_{\twg_4}}\lth})} \\
 &\equiv & T+\delta.
\eeq
Thus we conclude that $\WG_4(t)=1$ in the desired interval.

To prove part (b), observe that $F_{\ten_1}(t)=\WG_4(t)$ for all $t$, from\rf{eq-wgen},
and recall that $\en_1(0)=1$. From part (a), $F_{\ten_1}(t)=1$ for 
$t\in(\frac{1}{\alpha_{\tWG_4}}\ltho,T+\delta)$. 
On the other hand, $\en_1$ can  only switch from 1 to 0 at 
$t=\alpha_{\ten_1}^{-1}\lth$ which is larger than $\alpha_{\tWG_4}^{-1}\ltho$. So, in fact,
$\en_1(t)=1$ for all $0\leq t< T+\delta$.

Part (c) follows immediately by integration of the $\cEN_1$ equation.

To prove part (d), first recall $F_{\tci_1}=$ not $\EN_1$ and the initial conditions
$\ci_1(0)=0=\CI_1(0)=\CIA_1(0)=\CIR_1(0)$.
Therefore $\cci_1(t)$ increases up to $t=\frac{1}{\alpha_{\tEN_1}}\ltho$ and then decreases in
$\alpha_{\tEN_1}^{-1}\ltho<t<T+\delta$. Now note that the discrete variable $\ci_1(t)$ remains 0
in the whole interval $[0,T+\delta)$. This is because $\cci_1$ never reaches the $\theta$ threshold:
this would be attained at some $t\geq\alpha_{\tci_1}^{-1}\ltho$ but, since 
$\alpha_{\tci_1}^{-1}\ltho>\alpha_{\tEN_1}^{-1}\ltho$, the function $\cci_1$ starts decreasing 
before it could reach the value $\theta$.
Finally, from the rules of the Cubitus proteins it is immediate to see that 
$\CI_1(t)=\CIA_1(t)=\CIR_1(t)=0$ for $t\in[0,T+\delta)$.

To prove part (e), recall that $F_{\thh_1}=\EN_1$ and not $\CIR_1$. From part (a), it follows
that $F_{\thh_1}(t)=0$ in the interval $[0,\alpha_{\tEN_1}^{-1}\ltho)$ and 
$F_{\thh_1}(t)=1$ in the interval $(\alpha_{\tEN_1}^{-1}\ltho,T+\delta)$. Since $\hh_1(0)=1$, $\chh_1(t)$
decreases in the interval $[0,\alpha_{\tEN_1}^{-1}\ltho)$ but increases in 
$(\alpha_{\tEN_1}^{-1}\ltho,T+\delta)$.
The discrete value is $\hh_1(t)=1$ in the whole interval, since $\chh_1(t)$ remains above the
$\theta$ threshold. (The justification is similar to the case of $\ci_1(t)$ in part (d).) 

To prove part (f), note that part (e) and then the use of\rf{eq-ciCI}, allows us to simplify 
$F_{\tCIA_4}$:
\beq
  F_{\tCIA_4}(t)=\CI_4(t)\ \mbox{ and }\ \hh_1(t) = 1,\ \ \ t\in(\frac{1}{\alpha_{\tCI_4}}\ltho,T+\delta). 
\eeq
Thus
\beq
  \cCIA_4(t)=\left\{
        \begin{array}{ll}
           0, & 0\leq t\leq \frac{1}{\alpha_{\tCI_4}}\ltho \\
           1-e^{-\alpha_{\tCIA_4}\left(t-\frac{1}{\alpha_{\tCI_4}}\ltho\right)}, & 
               \frac{1}{\alpha_{\tCI_4}}\ltho<t\leq T+\delta,
        \end{array}
        \right.
\eeq 
and $\CIA_4(t)=1$ for $t\in[\frac{1}{\alpha_{\tCI_4}}\ltho+\frac{1}{\alpha_{\tCIA_4}}\ltho,T+\delta)$.
Observe that this interval is indeed nonempty, by assumption\rf{eq-separation}.
Finally, $F_{\tCIR_4}(t)=\CI_4(t)\mbox{ and not}\hh_1(t) = 0$, and hence $\CIR_4(t)=0$ for 
$t\in[0,T+\delta)$.

To prove part (g), we note that (from part (f))
\beq
  F_{\twg_4}(t)=1,\ \ \ t\in(\frac{1}{\alpha_{\tCI_4}}\ltho
                            +\frac{1}{\alpha_{\tCIA_4}}\ltho,T+\delta),
\eeq 
implying that $\cwg_4(t)$ increases in this interval. On the other hand, we know that
$\cwg_4(t)\geq\theta$ and $\wg_4(t)=1$ up to at least 
$t=\frac{1}{\alpha_{\twg_4}}\lth>\frac{1}{\alpha_{\tCI_4}}\ltho+\frac{1}{\alpha_{\tCIA_4}}\ltho$.
This shows that in fact $\wg_4(t)=1$ for all $t\in[0,T+\delta)$.
\bbox

{\it Proof of Theorem~\ref{th-glass2}:}
Since $\wg_4(0)=1$, from equations\rf{eq-hybrid-switch1},\rf{eq-hybrid-switch0}, 
we know that the earliest possible switching time from 1 to 0 is $\alpha_{\twg_4}^{-1}\lth$.
Applying Lemma~\ref{lm-wg4} with $T=\alpha_{\twg_4}^{-1}\lth$ establishes 
that $\wg_4(t)=1$ for $t\in[0,T+\delta)$, with $\delta$ given by\rf{eq-delta}.
Next, applying Lemma~\ref{lm-wg4} with $T=\alpha_{\twg_4}^{-1}\lth + k\delta$, $k\in\N$,
shows that $\wg_4(t)=1$ for $t\in[0,T+(k+1)\delta)$. Since $\delta$ is finite, we can conclude
by induction that $\wg_4(t)=1$ for all $t\geq0$.

To prove that $\PTC_1(t)\equiv0$, note that $\CIA_1(t)\equiv0$ (Lemma~\ref{lm-wg4}, 
with $T=+\infty$) implies $\ptc_1(t)\equiv0$. Since $\PTC_1(0)=0$ and $\PTC_1$ cannot
become expressed unless $\ptc_1$ is first expressed, the desired result follows.
\bbox

\section{Conclusions}

We discussed two alternative methods for modeling gene expression networks:
purely discrete Boolean methods and piecewise linear differential systems, which
combine continuous degradation with discrete synthesis.
For both methods we introduced new techniques for a deeper analysis of the networks 
with respect to perturbations in the timescales of the system. 
For the piecewise linear system we also studied the effect of the ON concentration thresholds.

We find that unrestricted variability in the duration of the diverse processes present 
in the network may lead to significant deviations from experimentally observed results,
thus suggesting the fragility of the developmental process under severe 
perturbations (the asynchronous
algorithm fails to predict the correct pattern with probability 50\%, and the Glass-type  
system fails with probability at least 10\%).

Another set of numerical experiments introduces a separation between timescales of 
post-translational and transcription/translation processes, and in practice 
``updates mRNAs later than proteins''.
In this context, the piecewise linear Glass-type system indicates a remarkable robustness 
of the Boolean model in predicting the final gene expression pattern.
Indeed, we provide a theoretical proof that the Glass-type system always correctly
generates the wild type development (i.e. the convergence to the 
wild type steady state when started from the wild type initial state),
under the separation of timescales assumption,
and appropriate OFF/ON thresholds.
The asynchronous model's predictions depend very much on the degree of separation between
timescales. As the intervals $A_{\tProt}$ and $A_{\tmRNA}$ become closer, the asynchronous 
algorithm increasingly fails to generate the wild type pattern.
This leads us to conclude that a strong separation between timescales of 
post-translational and transcription/translation processes is necessary for establishing
the regular gene expression pattern in the segment polarity network.
From analysis of the piecewise linear model we conclude that the fraction of maximal 
concentration above which a protein or mRNA is effectively ON needs to be quite small,
below 50\%. Higher concentration thresholds may disrupt the development process.

The comparison between discrete and continuous models shows clearly that sudden
transitions may happen in discrete systems and lead to a false result; such
sudden transitions are smoothed out in continuous models which prevent generation
of false results (see also~\cite{m05}).
Nevertheless, we conclude that both models agree in predicting the fundamental sequences of 
gene expression that irreversibly lead to a deviation in the development towards a 
mutant state.

By combining continuous-time techniques 
with discrete events, we can with great generality 
explore and sample the space of all possible timescales 
as well as of effective ON levels.
Moreover, as information about the mRNA/protein lifetimes, decay rates or
activation thresholds becomes available, it can be straightforwardly incorporated
by fixing the corresponding inverse scaling factor $\alpha_i^{-1}$. The hybrid model
retains the ease of Boolean models in determining the steady states
corresponding to gene knockouts and perturbed initial conditions. 
It is straightforward to calculate which mutant patterns result from each gene knockout.
Here we also studied the the effect of perturbations on the prepattern, by simulating delay 
in initial expression of each gene. We find that the system is vulnerable to large delays
(larger than two time units) in expression of any gene -- except for $\ci$ --
and,  in such delayed conditions, the mutant state characteristic to that gene knockout is
generated.
For low order delays, the system typically recovers and proceeds through the correct wild 
type development.

The Glass-type system 
with time separation, as a model of the segment polarity gene network,
reflects the conclusion of von Dassow et al. that the topology of the
network is more important than the fine-tuning of the kinetic parameters
\cite{dmmo00}, since its results are robust for a large region
of parameter (scaling factor, 
activation threshold)
space. Due to its underlying Boolean structure,
the model also intrinsically incorporates the
recent finding of Ingolia that parameter sets need to 
satisfy certain constraints - that ensure the bistability of certain
genes- to lead to correct solutions\cite{i04}. Taken together, the results of the 
synchronous~\cite{ao03}, asynchronous~\cite{cas05} Boolean and hybrid models
convincingly demonstrate the Boolean 
models' capability for effectively describing the basic structure and functioning of 
gene control networks when detailed kinetic information is unavailable.

\bibliography{drosophila}

\begin{thebibliography}{10}

\bibitem{giot03}
L.~Giot, J.~S. Bader, C.~Brouwer, A.~Chaudhuri, and B.~Kuang et~al.
\newblock A protein interaction map of drosophila melanogaster.
\newblock {\em Science}, 302:1727--1736, 2003.

\bibitem{han04}
J.~D. Han, N.~Bertin, T.~Hao, D.~S. Goldberg, and G.~F.~Berriz et~al.
\newblock Evidence for dynamically organized modularity in the yeast
  protein-protein interaction network.
\newblock {\em Nature}, 430:88--93, 2004.

\bibitem{li04}
S.~Li, C.~M. Armstrong, N.~Bertin, H.~Ge, and et~al. S.~Milstein.
\newblock A map of the interactome network of the metazoan c. elegans.
\newblock {\em Science}, 303:540--543, 2004.

\bibitem{lee02}
T.~I. Lee, N.~J. Rinaldi, F.~Robert, D.~T. Odom, and et~al. Z.~Bar-Joseph.
\newblock Transcriptional regulatory networks in saccharomyces cerevisiae.
\newblock {\em Science}, 298:799--804, 2002.

\bibitem{a05}
R.~Albert.
\newblock Scale-free networks in cell biology.
\newblock {\em Journal of cell science}, 118:4947--4957, 2005.

\bibitem{lbyst04}
N.~M. Luscombe, M.~M. Babu, H.~Y. Yu, M.~Snyder, S.~S. Teichmann, and
  M.~Gerstein.
\newblock Genomic analysis of regulatory network dynamics reveals large
  topological changes.
\newblock {\em Nature}, 431:308--312, 2004.

\bibitem{st01}
L.~S\'anchez and D.~Thieffry.
\newblock A logical analysis of the {\it drosophila} gap-gene system.
\newblock {\em J. Theor. Biol.}, 211:115--141, 2001.

\bibitem{ao03}
R.~Albert and H.~G. Othmer.
\newblock The topology of the regulatory interactions predicts the expression
  pattern of the {\it drosophila} segment polarity genes.
\newblock {\em J. Theor. Biol.}, 223:1--18, 2003.

\bibitem{mta99}
L.~Mendoza, D.~Thieffry, and E.~R. Alvarez-Buylla.
\newblock Genetic control of flower morphogenesis in {\it arabidopsis
  thaliana}: a logical analysis.
\newblock {\em Bioinformatics}, 15:593--606, 1999.

\bibitem{epa04}
C.~Espinosa-Soto, P~Padilla-Longoria, and E.~R. Alvarez-Buylla.
\newblock A gene regulatory network model for cell-fate determination during
  arabidopsis thaliana flower development that is robust and recovers
  experimental gene expression profiles.
\newblock {\em Plant Cell}, 16:2923--2939, 2004.

\bibitem{dmmo00}
G.~von Dassow, E.~Meir, E.M. Munro, and G.M. Odell.
\newblock The segment polarity network is a robust developmental module.
\newblock {\em Nature}, 406:188--192, 2000.

\bibitem{cas05}
M.~Chaves, R.~Albert, and E.D. Sontag.
\newblock Robustness and fragility of boolean models for genetic regulatory
  networks.
\newblock {\em J. Theor. Biol.}, 235:431--449, 2005.

\bibitem{g75}
L.~Glass.
\newblock Classification of biological networks by their qualitative dynamics.
\newblock {\em J. Theor. Biol.}, 54:85--107, 1975.

\bibitem{eg00}
R.~Edwards and L.~Glass.
\newblock Combinatorial explosion in model gene networks.
\newblock {\em Chaos}, 10:691--704, 2000.

\bibitem{gk73}
L.~Glass and S.A. Kauffman.
\newblock The logical analysis of continuous, nonlinear biochemical control
  networks.
\newblock {\em J. Theor. Biol.}, 39:103--129, 1973.

\bibitem{hs92}
J.~E. Hooper and M.P. Scott.
\newblock The molecular genetic basis of positional information in insect
  segments.
\newblock In W.~Hennig, editor, {\em Early Embryonic Development of Animals},
  pages 1--49. Springer, Berlin, 1992.

\bibitem{gakt00}
A.~Gallet, C.~Angelats, S.~Kerridge, and P.P. Th\'erond.
\newblock Cubitus interruptus-independent transduction of the hedgehog signal
  in {\it drosophila}.
\newblock {\em Development}, 127:5509--5522, 2000.

\bibitem{tek92}
T.~Tabata, S.~Eaton, and T.B. Kornberg.
\newblock The {\it drosophila hedgehog} gene is expressed specifically in
  posterior compartment cells and is a target of {\it engrailed} regulation.
\newblock {\em Genes \& Dev.}, 6:2635--2645, 1992.

\bibitem{cgg94}
K.M. Cadigan, U.~Grossniklaus, and W.~J. Gehring.
\newblock Localized expression of {\it sloppy paired} protein maintains the
  polarity of {\it drosophila} parasegments.
\newblock {\em Genes \& Dev.}, 8:899--913, 1994.

\bibitem{av03}
C.~Alexandre and J.~P. Vincent.
\newblock Requirements for transcriptional repression and activation by
  engrailed in {\it drosophila} embryos.
\newblock {\em Development}, 130:729--739, 2003.

\bibitem{sg04}
D.~Swantek and J.~P. Gergen.
\newblock Ftz modulates runt-dependent activation and repression of segment
  -polarity gene transcription.
\newblock {\em Development}, 131:2281--2290, 2004.

\bibitem{tsi}
D.P. Bertsekas and J.N. Tsitsiklis.
\newblock {\em Parallel and Distributed Computation, Numerical Method}.
\newblock Prentice Hall, Englewood Cliffs, New Jersey, 1989.

\bibitem{jghpsg04}
H.~de~Jong, J.L. Gouz\'e, C.~Hernandez, M.~Page, T.~Sari, and J.~Geiselmann.
\newblock Qualitative simulation of genetic regulatory networks using piecewise
  linear models.
\newblock {\em Bull. Math. Biol.}, 66:301--340, 2004.

\bibitem{gedeon}
T.~Gedeon.
\newblock Attractors in continuous-time switching networks.
\newblock {\em Communications on Pure and Applied Analysis}, 2:187--209, 2003.

\bibitem{m05}
A.~Mochizuki.
\newblock An analytical study of the number of steady states in gene regulatory
  networks.
\newblock {\em J. Theor. Biol.}, 236:291--310, 2005.

\bibitem{i04}
N.T. Ingolia.
\newblock Topology and robustness in the drosophila segment polarity network.
\newblock {\em PLoS Biology}, 2:0805--0815, 2004.

\end{thebibliography}

\clearpage

\clearpage

\clearpage

\clearpage

\clearpage

\clearpage

\clearpage

\clearpage

\end{document}